\documentclass[twocolumn,aps,prl,superscriptaddress]{revtex4-2}

\usepackage{amsmath}
\usepackage{graphicx}
\usepackage{MnSymbol}
\usepackage{parskip}
\usepackage{makecell}
\usepackage{multirow}
\usepackage{siunitx}
\begin{document}

\title{Rubber wear on concrete: dry and in-water conditions}

\author{R. Xu}
\affiliation{Peter Gr\"unberg Institute (PGI-1), Forschungszentrum J\"ulich, 52425, J\"ulich, Germany}
\affiliation{State Key Laboratory of Solid Lubrication, Lanzhou Institute of Chemical Physics, Chinese Academy of Sciences, 730000 Lanzhou, China}
\affiliation{MultiscaleConsulting, Wolfshovener str. 2, 52428 J\"ulich, Germany}

\author{N. Miyashita}
\affiliation{The Yokohama Rubber Company, 2-1 Oiwake, Hiratsuka, Kanagawa 254-8601, Japan}

\author{B.N.J. Persson}
\affiliation{Peter Gr\"unberg Institute (PGI-1), Forschungszentrum J\"ulich, 52425, J\"ulich, Germany}
\affiliation{State Key Laboratory of Solid Lubrication, Lanzhou Institute of Chemical Physics, Chinese Academy of Sciences, 730000 Lanzhou, China}
\affiliation{MultiscaleConsulting, Wolfshovener str. 2, 52428 J\"ulich, Germany}

\begin{abstract}
{\bf Abstract}: 
Rubber wear results from the removal of small (micrometer-sized) rubber particles through crack propagation. In this study, we investigate the wear behavior of Styrene-Butadiene Rubber (SBR) and Natural Rubber (NR) sliding on two different concrete surfaces under dry and wet (in water) conditions. Experiments are conducted at low sliding speeds ($\approx 3 \ {\rm mm/s}$) to minimize frictional heating and hydrodynamic effects. For two SBR compounds, we observe significantly higher wear rates in water compared to the dry state, with enhancement factors of $1.5-2.5$ for a low-glass-transition-temperature SBR compound ($T_{\rm g} = -50^\circ {\rm C}$) and approximately $4$ for a higher-glass-transition compound ($T_{\rm g} = -7^\circ {\rm C}$). In contrast, the NR compound showed no wear in water at low nominal contact pressures ($\sigma_0 \approx 0.12$, $0.16$, and $0.25 \ {\rm MPa}$), while at higher pressures ($\sigma_0 \approx 0.36$ and $0.49 \ {\rm MPa}$), the wear rates in dry and in-water states are similar. The findings provide insights into the mechanisms of rubber wear under varying environmental and 
mechanical conditions, highlighting the influence of material properties, interfacial effects, and applied pressures on wear behavior.
\end{abstract}

\maketitle

\setcounter{page}{1}
\pagenumbering{arabic}




{\bf Corresponding author:} B.N.J. Persson, email: b.persson@fz-juelich.de
\vskip 0.3cm

{\bf 1 Introduction}

Wear is the progressive loss of material from a solid body due to its contact and 
relative movement against a surface  \cite{ToBe,wear1,wear2,Rabi1, Rabi2, Moli1, Moli2, Roland,Moli,W1,W2}. 
Wear particles can adversely impact the health of living organisms and lead to 
the breakdown of mechanical devices. There are several limiting wear processes, 
commonly categorized as {\it fatigue wear}, {\it abrasive wear}, and {\it adhesive wear}.
In this study, we focus on fatigue wear, which occurs when a rubber block slides on a rigid countersurface 
with ``smooth roughness.'' In this scenario, the stress concentrations in asperity contact regions are relatively low, 
necessitating multiple contacts to remove polymer particles. This wear process involves fatigue failure rather 
than tensile failure, where material removal occurs gradually. The abrasion caused by this failure mode is referred to as fatigue wear.

Understanding crack propagation is critical for analyzing rubber wear. 
The crack or tearing energy $\gamma$ (defined as the energy per unit area required to separate surfaces at a crack tip) 
is a key measure of material resistance to crack growth  \cite{Paris}. 
For rubber, $\gamma$ can range from $\sim 10^2$ to $\sim 10^5 \ {\rm J/m^2}$, depending on factors such as 
crack tip velocity and temperature. By comparison, the crack energy for brittle crystalline solids is on the order of 
$\sim 1 \ {\rm J/m^2}$, even for materials with strong covalent bonds like diamonds. The large $\gamma$ values in polymers 
arise from energy contributions due to chain stretching, uncrosslinked chain pull-out, and mechanisms like crazing, cavitation, 
and viscoelastic dissipation near the crack tip.

The crack energy $\gamma$ has been extensively studied for cases of constant crack tip velocity  \cite{Gent} 
and oscillating strains  \cite{Rivlin, NatRub, Ghosh}, yielding similar results in both cases. 
Under oscillatory strain, the crack tip displacement $\Delta x$ per strain cycle depends on $\gamma$. 
Below a lower critical value $\gamma_0$ (typically $\sim 10^2 \ {\rm J/m^2}$ for rubber), 
no crack growth occurs (in vacuum). As $\gamma$ approaches the ultimate tear strength $\gamma_2$ 
(typically $\sim 10^5 \ {\rm J/m^2}$ for rubber), $\Delta x$ diverges. However, unless $\gamma$ is close to $\gamma_2$, 
the crack tip displacement $\Delta x$ is very small. Therefore, multiple stress cycles 
may be required to remove a particle from a rubber surface under {\it fatigue wear}. 
In normal atmospheric conditions, very slow crack propagation can occur for $\gamma < \gamma_0$ due to
stress corrosion, where the stretched bonds at the crack tip are broken by reaction with foreign molecules, e.g.,
oxygen or ozone.

Crack energy measurements are typically conducted on macroscopic samples 
(linear dimensions $\sim 1 \ {\rm cm}$). These measurements may not fully represent the smaller length scales 
relevant to polymer wear, where particles as small as $\sim 1 \ {\rm \mu m}$ are removed. At these scales, contributions 
from mechanisms such as viscoelastic energy dissipation or cavitation may be reduced. 
Additionally, in sliding, the asperity-induced deformation spans a broad frequency range 
($\omega \approx v/r_0$, where $v$ is the sliding speed and $r_0$ is the contact size), 
in contrast to the fixed frequency conditions used in conventional tearing energy experiments.

In Ref.  \cite{ToBe}, a theoretical framework was developed to describe sliding wear in rubber. 
This theory incorporates the complex interplay between crack propagation, material properties, 
and contact mechanics. Building on this foundation, the current study presents additional experimental data 
to test the validity of the theory and explore new aspects of rubber wear.

We use two styrene-butadiene rubber (SBR) compounds, one with silica filler and one with carbon black filler,
and a natural rubber (NR) compound with carbon filler. Two concrete blocks with different surface roughness are used as substrates, enabling the investigation of the role of surface characteristics in determining wear behavior. Environmental conditions, such as dryness or wetness, are also an important factor in rubber wear. 
In certain regions such as Northern Europe, where approximately one-third of the days in a year experience precipitation, the wear of tires under wet conditions becomes significant. In other applications, such as dynamic rubber seals operating in water or conveyor belts transporting wet materials, the rubber may remain 
in contact with wet surfaces for extended periods. We therefore also investigate the wear of two compounds on dry substrate and in water, thereby providing a comprehensive understanding of rubber wear mechanisms under various conditions.

\vskip 0.3cm
{\bf 2 Experimental methods}

\begin{figure}[h]
\includegraphics[width=0.95\columnwidth]{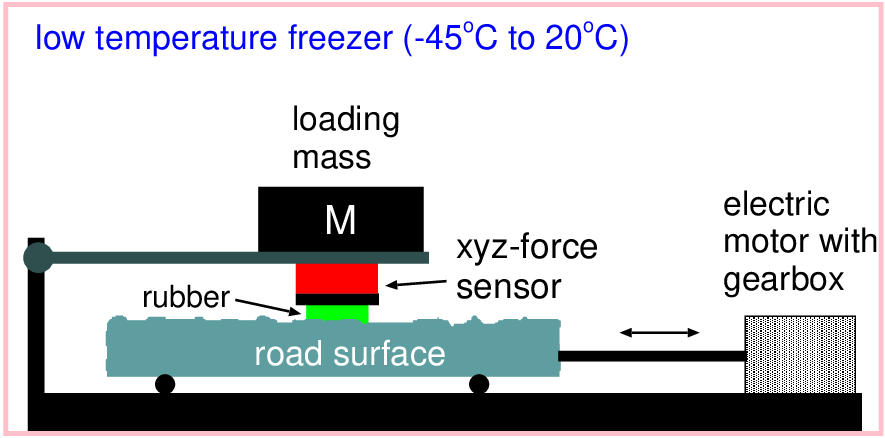}
\caption{\label{LowTemperaturePic.eps}
Schematic of the friction instrument used in the present study, allowing for linear reciprocal motion at constant speed.}
\end{figure}

Fig. \ref{LowTemperaturePic.eps} schematically illustrates the
linear friction tester used in most of the wear studies presented below. All measurements in the present study are conducted at room temperature. 

The setup consists of rubber blocks with the nominal contact area $A_0 \approx 7  \ {\rm cm^2}$ or $\approx 16 \ {\rm cm^2}$ 
glued onto an aluminum plate sample holder, which is connected to a force cell (depicted as a red box in the figure). 
The rubber specimen can move vertically with the carriage to adapt to the substrate profile. The normal load is adjustable by adding additional steel weights on top of the force cell.

The substrate (in this case, a concrete block) is attached to the machine table, which moves transversal using a servo drive via a gearbox. This setup allows precise control of the relative velocity between the rubber specimen and the substrate sample, while the force cell records data on the normal force and friction force. 

After the metal plate with the rubber blocks is slid the same distance forward and backward and returns to the starting position, the wear rate is determined by measuring the mass difference of the rubber blocks and metal plate before and after sliding. A high-precision balance (Mettler Toledo analytical balance, model MS104TS/00) with a sensitivity of $0.1 \ {\rm mg}$ is used for this purpose. After each sliding cycle, the surface is cleaned using a brush or a single-use nonwoven fabric.

The viscoelastic modulus in the linear response regime and the non-linear stress-strain curves for the two rubber compounds are given in Appendix A.

The surface roughness of the two concrete surfaces used in this study was characterized using a Mitutoyo Portable Surface Roughness Measurement Surftest SJ-410. The rough and smooth surfaces exhibit root-mean-square (rms) roughness amplitudes of $51.3$ and $33.6 \ {\rm \mu m}$, respectively, and rms slopes of $0.46$ and $0.37$. Further details on the measurement procedures and the corresponding surface power spectra are provided in Appendix B.

\vskip 0.3cm
{\bf 3 Experimental results}

All measurements are performed at room temperature $T=20^\circ {\rm C}$ at a sliding speed of $3 \ {\rm mm/s}$, the low sliding velocity is chosen to avoid frictional heating. In most cases, each test consisted of multiple $20 \ {\rm cm}$ forward and backward sliding motions. The rubber mass density is assumed to be $\rho = 1.2 \ {\rm g/cm^3}$ when calculating the wear volume.

We measured the wear rate and the friction force for SBR on ``smooth" and ``rough" concrete surfaces, and NR on the ``smooth" concrete surface. We used a low $T_{\rm g}$ ($\approx -50^\circ {\rm C}$) SBR with silica (SBRs) or carbon black fillers (SBRc). 
The fillers constituted $20 \ {\rm vol}\%$ of the compounds. The two compounds had similar mechanical properties, 
except for the elongation at failure, which is $40\%$ higher for the carbon black-filled compound (factors $2.94$ and $4.10$, respectively).
We also performed studies for the NR compound used in Ref. \cite{ToBe}. The NR compound has $30 \ {\rm weight}\%$ carbon filler (NRc)
and the glass transition temperature $T_{\rm g} \approx -66^\circ {\rm C}$.

\begin{figure}[h]
	\includegraphics[width=0.47\textwidth,angle=0.0]{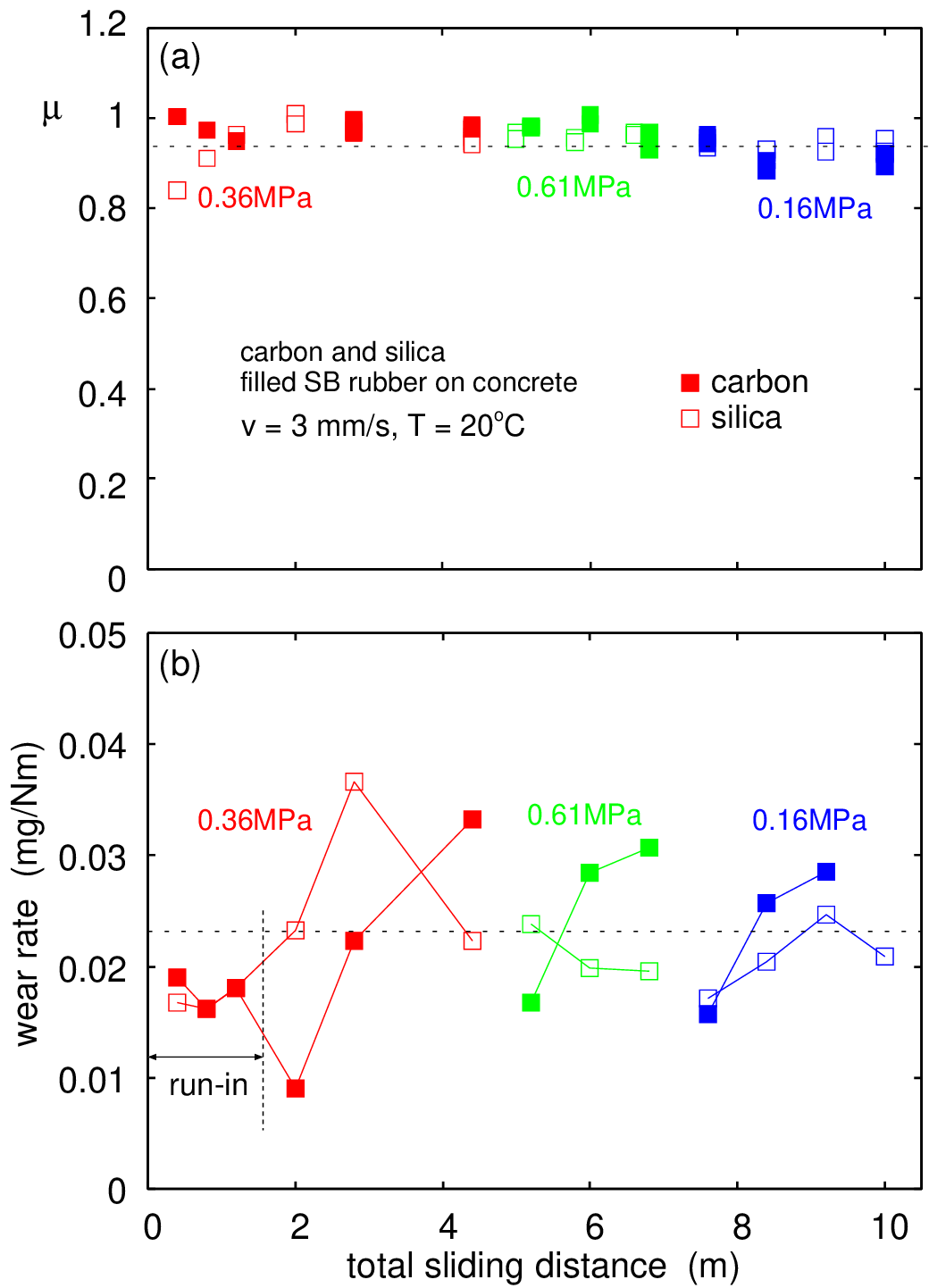}
\caption{(a) The friction coefficient $\mu$ and (b) the mass loss per unit sliding distance divided by the normal force $\Delta m/F_{\rm N}L$, as a function of the sliding distance on ``smooth'' concrete surfaces. The open and filled squares represent SBR with silica and carbon black fillers, respectively. Colors indicate different nominal contact pressures $\sigma_0$. The average friction coefficient is $\mu \approx 0.95$, and the average wear rate is 
$\Delta m/F_{\rm N}L \approx 0.023 \ {\rm mg/Nm}$, as indicated by the black dashed lines.}
\label{1s.2wearrate.all.loads.eps}
\end{figure}

\begin{figure}[h]
\includegraphics[width=0.45\textwidth,angle=0.0]{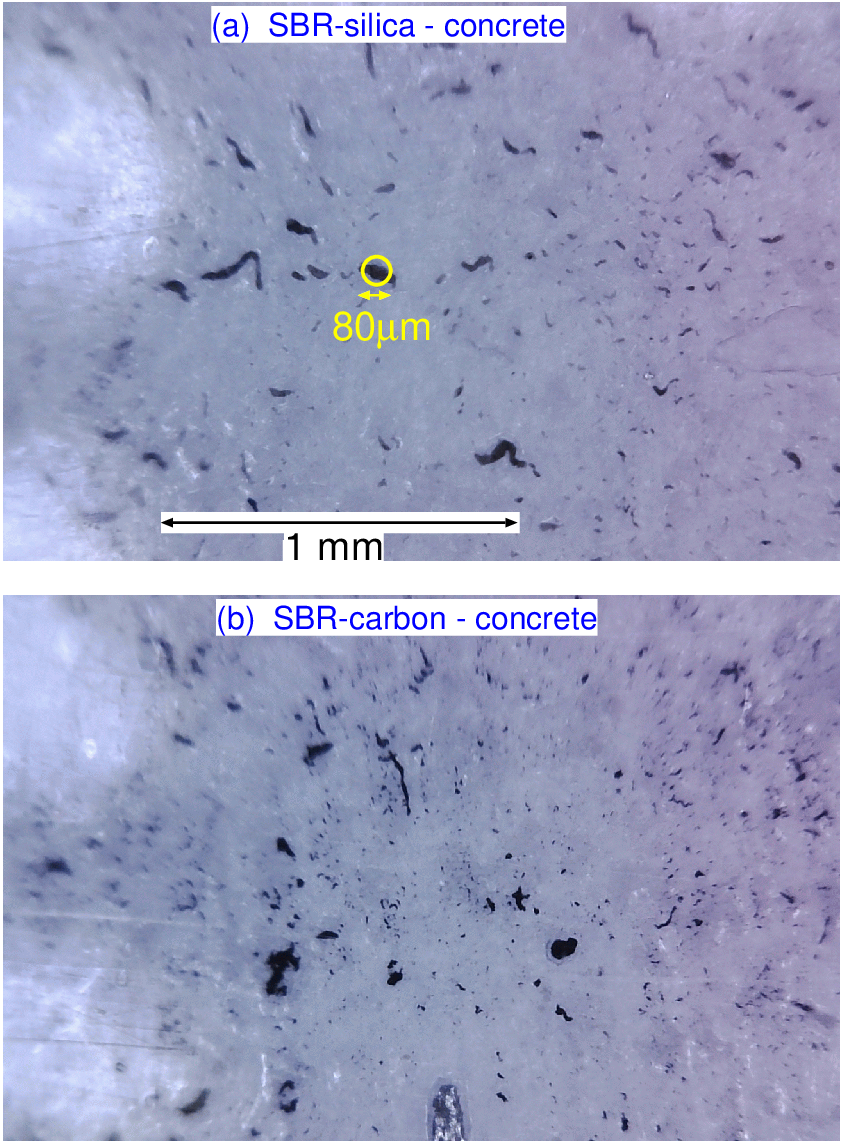}
\caption{\label{NewPicWearParticles.1.eps}
        Optical images of the wear particles on ``smooth'' concrete surfaces of (a) Silica-filled SBR compound and (b) carbon-filled SBR compound. In both cases, the sliding distance is $18.5 \ {\rm cm}$ and the nominal contact pressure is $\sigma_0 = 0.056 \ {\rm MPa}$, the measurements are conducted separately using the Leonardo da Vinci slider. 
}
\end{figure}

\begin{figure}[h]
\includegraphics[width=0.47\textwidth,angle=0.0]{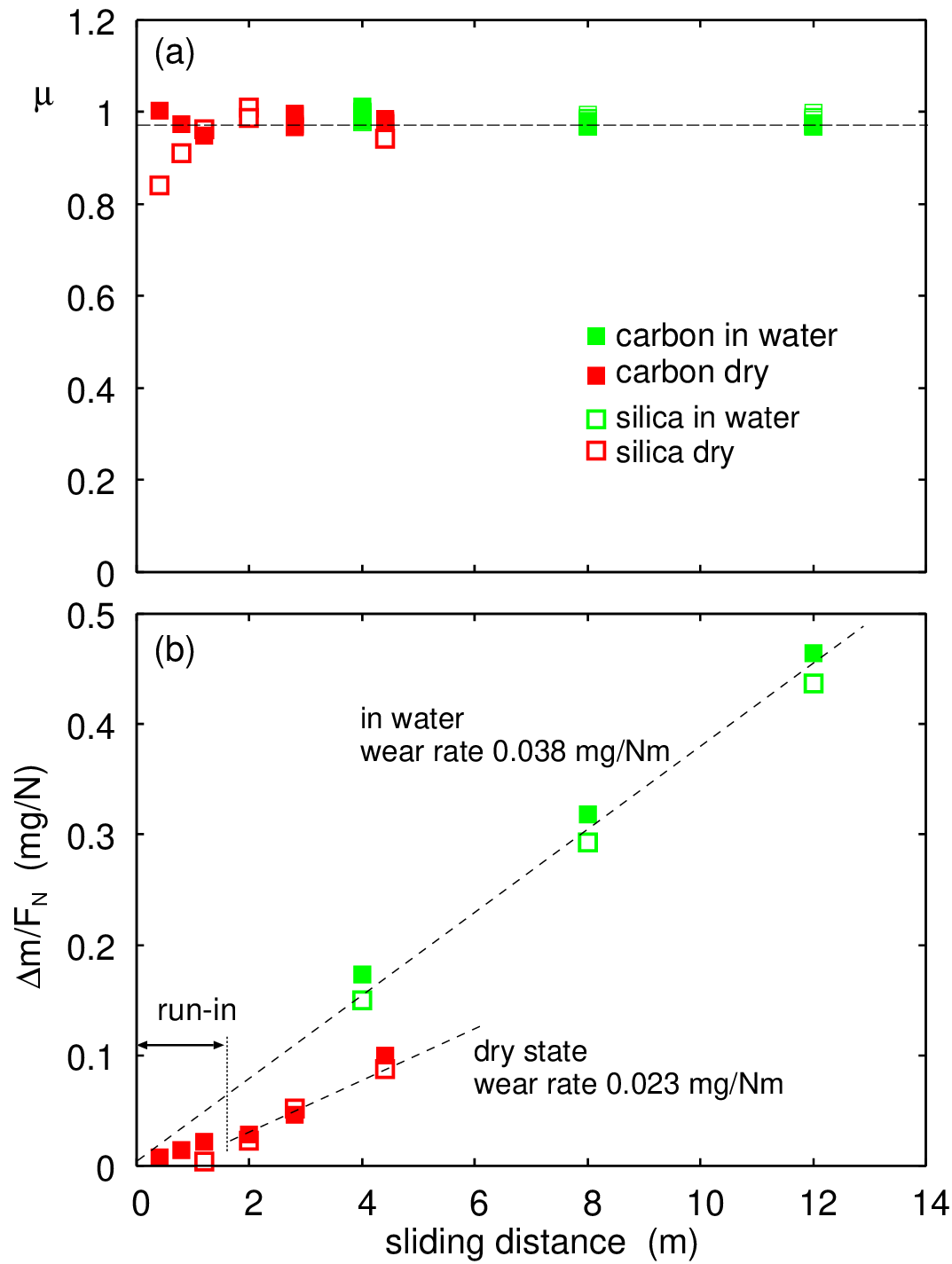}
\caption{(a) The friction coefficient $\mu$ and (b) the mass loss divided by the normal force, $\Delta m/F_{\rm N}$, for carbon- and silica-filled SBR blocks as a function of the sliding distance on ``smooth'' concrete surfaces at a nominal contact pressure of $0.36 \ {\rm MPa}$. The red and pink colors represent the dry and in-water state. The filled and open squares correspond to SBR with carbon and silica fillers, respectively. The dashed lines in (b) indicate wear rates of $\Delta m/F_{\rm N}L \approx 0.038 \ {\rm mg/Nm}$ in water and $0.023 \ {\rm mg/Nm}$ in the dry state, or $\Delta V/F_{\rm N}L \approx 0.032 \ {\rm mm^3/Nm}$ in water and $0.019 \ {\rm mm^3/Nm}$ in the dry state.}
\label{1s.2wearrate.WATER.eps}
\end{figure}

\begin{figure}[h]
\includegraphics[width=0.35\textwidth,angle=0.0]{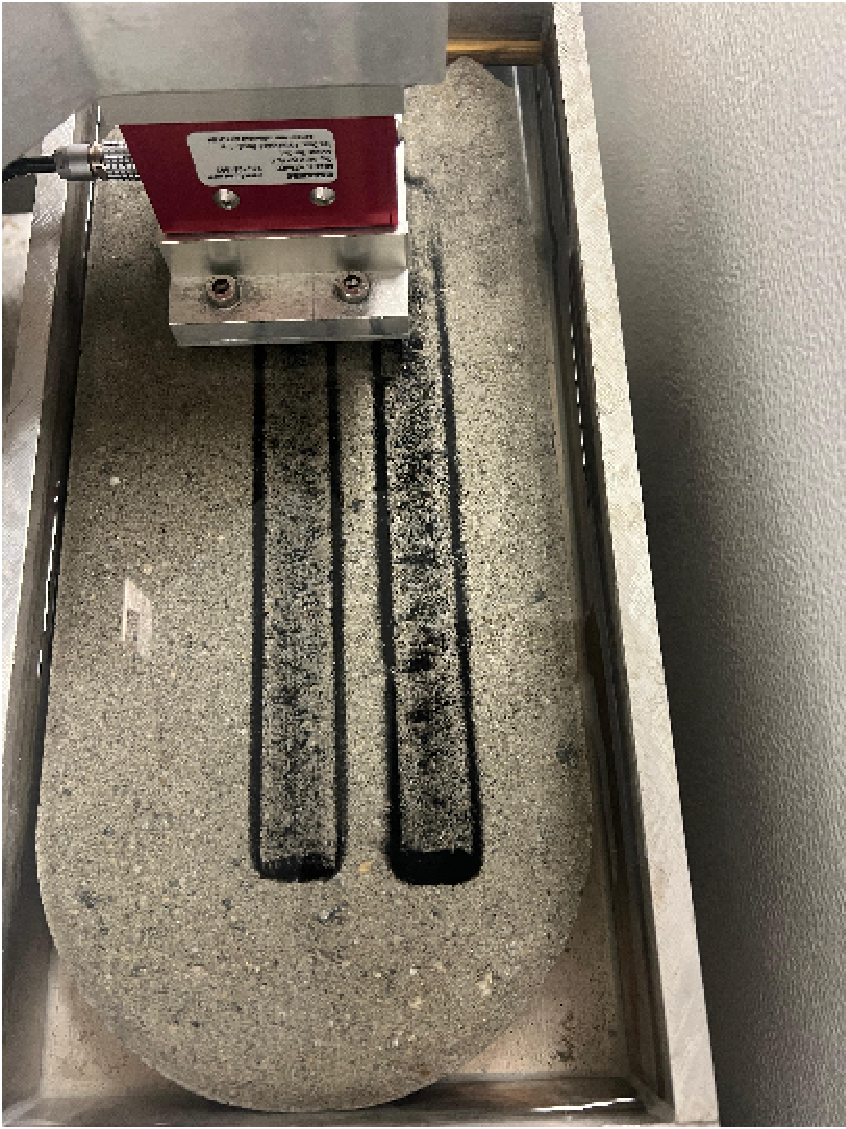}
\caption{\label{ALLwearCarbon.eps}
Wear tracks for the carbon-filled SBR block sliding on a ``smooth'' concrete surface in water.
The sliding motion consisted of 10 sliding cycles, each comprising a $20 \ {\rm cm}$ forward and backward motion.
}
\end{figure}

\begin{figure}[h]
\includegraphics[width=0.45\textwidth,angle=0.0]{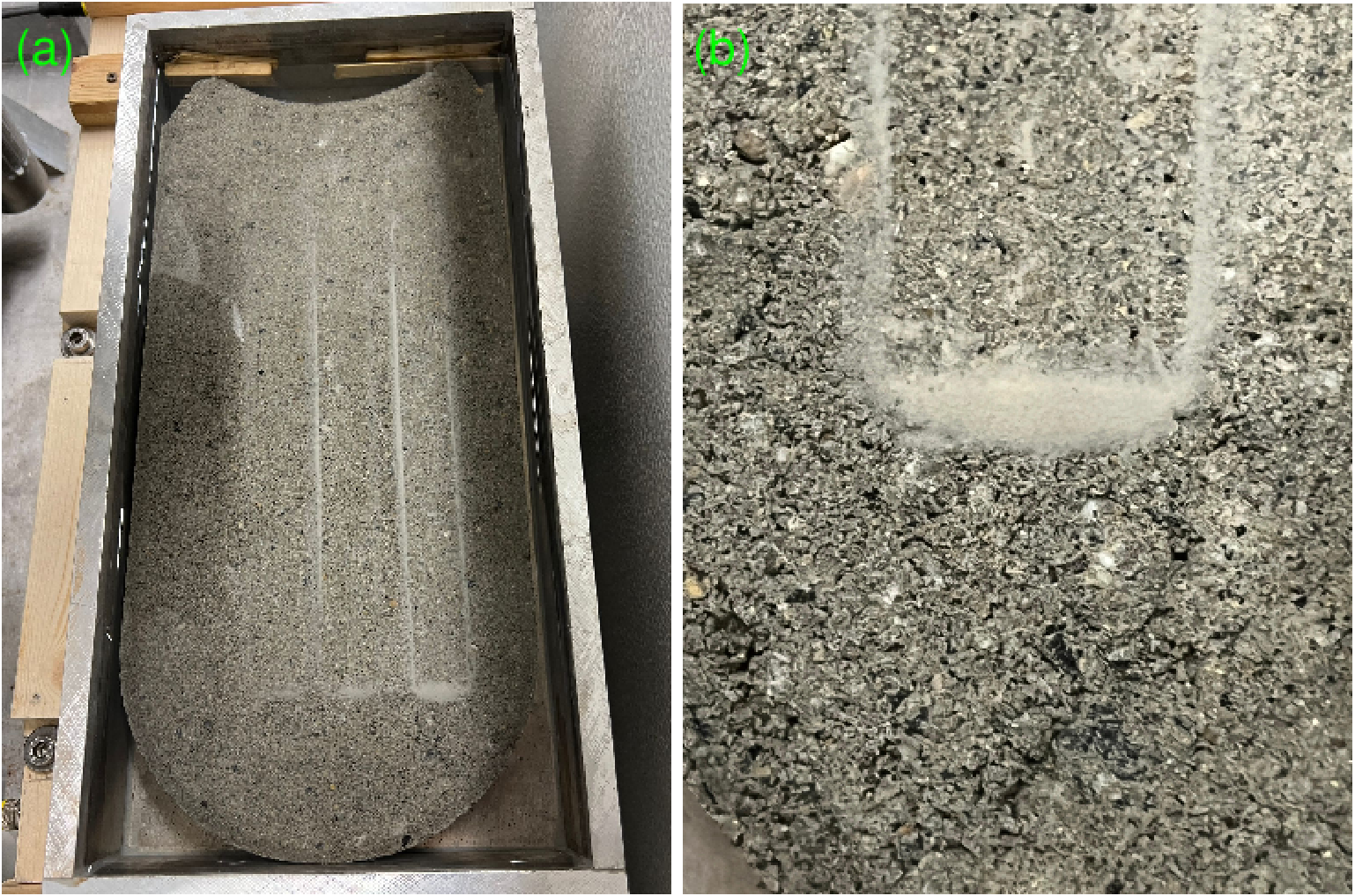}
\caption{\label{ALLwearSilicone.eps}
(a) Wear tracks for the silica-filled SBR block sliding on a ``smooth'' concrete surface in water.
The sliding motion consisted of 10 sliding cycles, each comprising a $20 \ {\rm cm}$ forward and backward motion.
(b) Magnified view of the wear particles accumulated at the edge of one sliding track.
}
\end{figure}

\begin{figure}[h]
\includegraphics[width=0.45\textwidth,angle=0.0]{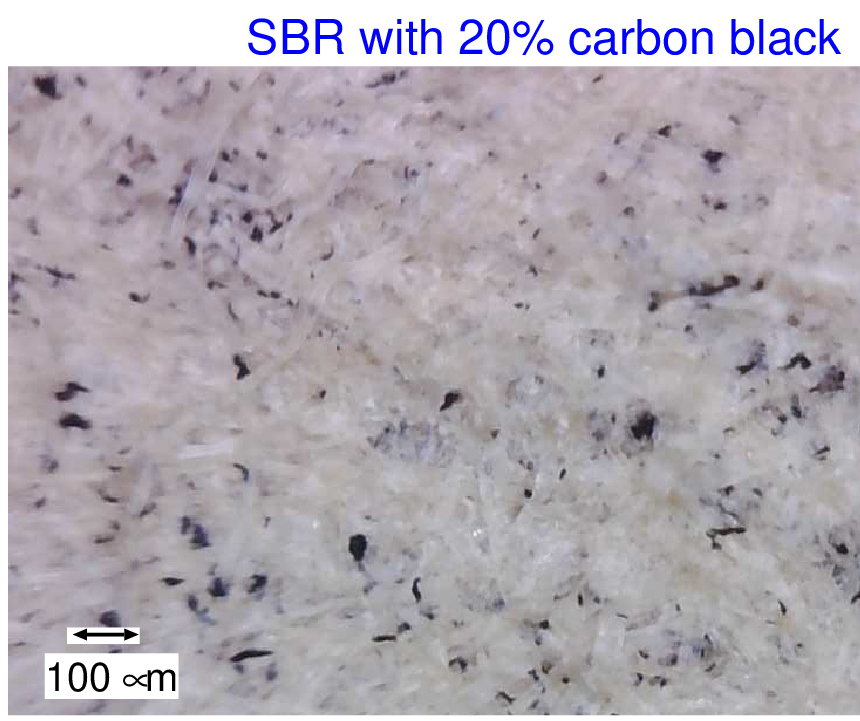}
\caption{\label{SBcPerticlesWater.ps}
Wear particles from the experiment with the carbon black-filled SBR compound sliding in water on a ``smooth''
concrete surface (Fig. \ref{ALLwearCarbon.eps}), collected on filter paper.
}
\end{figure}

\begin{figure}[h]
\includegraphics[width=0.45\textwidth,angle=0.0]{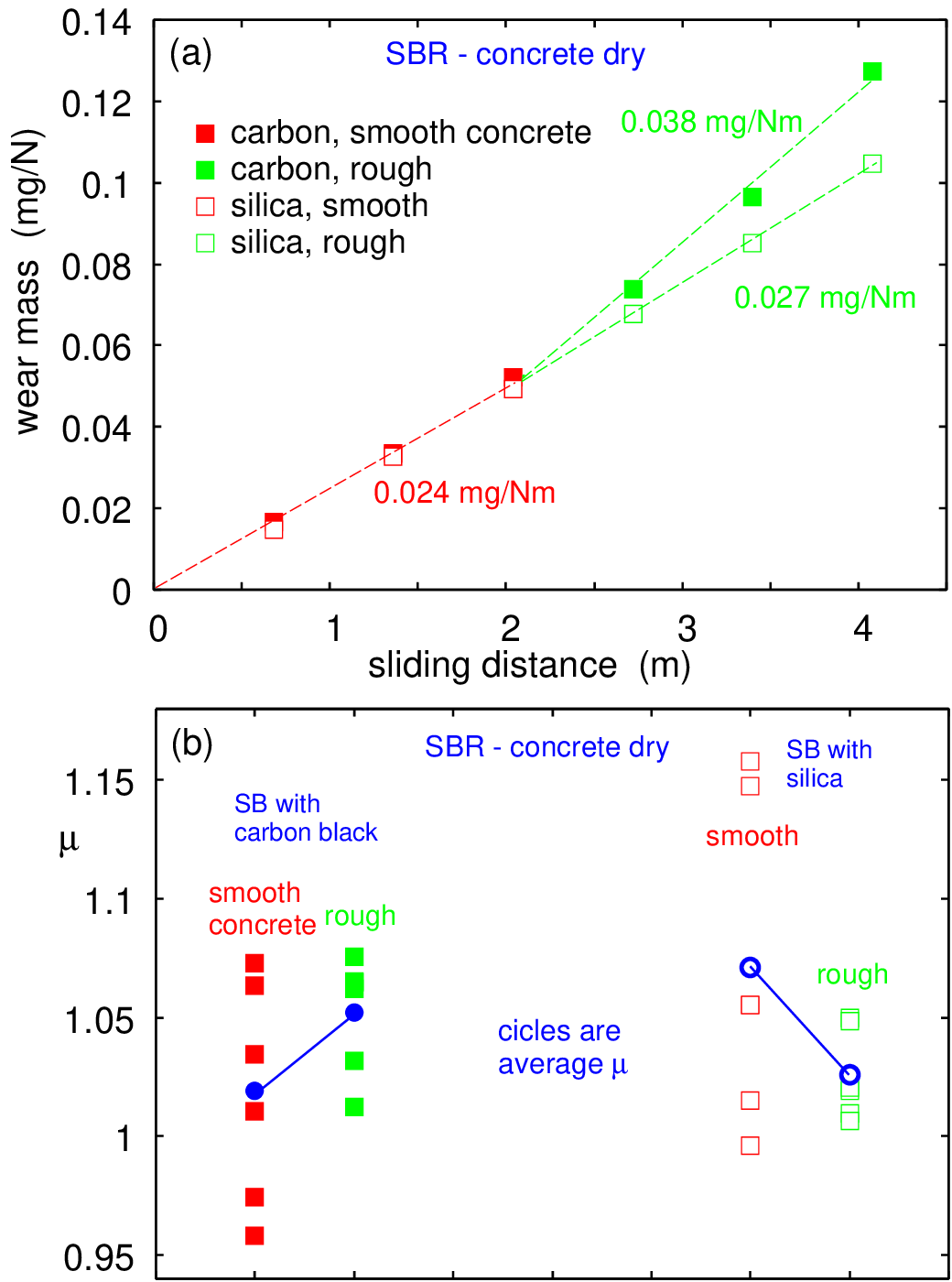}
\caption{\label{ZZ1distance.2mass.eps}
(a)Mass loss of rubber blocks as a function of sliding distance for SBR with carbon black filler (filled squares) 
and silica filler (open squares) on dry smooth (red) and rough (green) concrete surfaces with nominal contact pressure $\sigma_0 = 0.36 \ {\rm MPa}$. (b) corresponding friction coefficients of the average value (circles) and for each sliding cycle (squares). The average values are 1.019, 1.052, 1.071, and 1.026 
for the SBRc-smooth, SBRc-rough, SBRs-smooth, and SBRs-rough systems, respectively.
}
\end{figure}

\begin{figure}[h!]
\includegraphics[width=0.45\textwidth,angle=0.0]{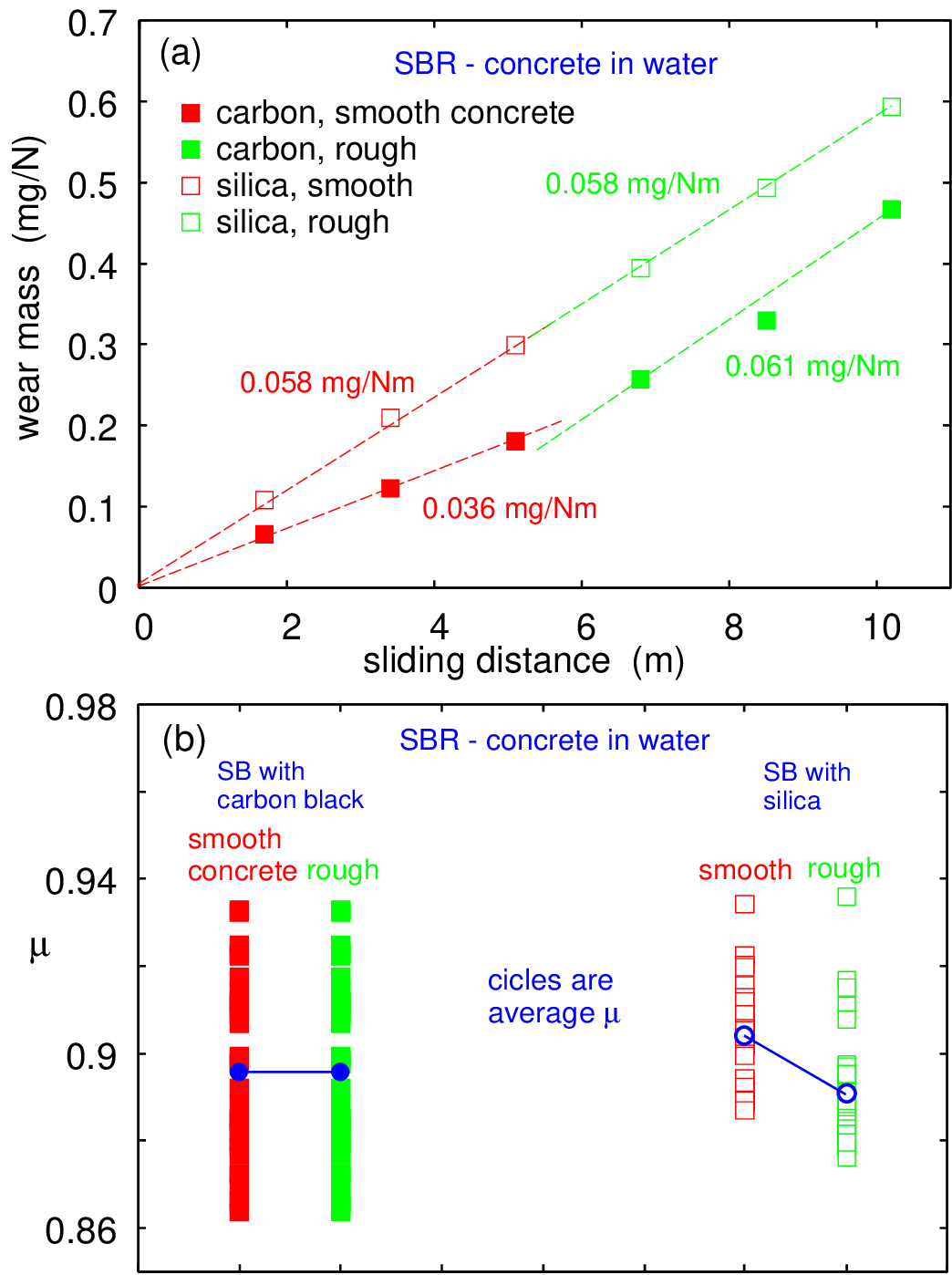}
\caption{\label{XX1distance.2wearmass.SB.wet.eps}
(a) Mass loss of rubber blocks as a function of sliding distance for SBR with carbon black filler (filled squares) and silica filler (open squares) on smooth (red) and rough (green) concrete surfaces in water with nominal contact pressure $\sigma_0 = 0.36 \ {\rm MPa}$. (b) corresponding friction coefficients of the average value (circles) and for each sliding cycle (squares). The average values are 0.896, 0.896, 0.904, and 0.891 for the SBRc-smooth, SBRc-rough, SBRs-smooth, and SBRs-rough systems, respectively.
}
\end{figure}

\vskip 0.2cm
{\bf SBR on concrete, dry and in water}

Experimental conditions and results for SBR are summarized in Table. \ref{SBR} and shown in Fig. \ref{1s.2wearrate.all.loads.eps}.
Fig. \ref{1s.2wearrate.all.loads.eps} (a) shows the friction coefficient $\mu$ and 
(b) the mass loss per unit sliding distance divided by the normal force $\Delta m/F_{\rm N}L$, 
as a function of the sliding distance on concrete surfaces. 
The open and filled squares represent SBRs and SBRc, respectively. 
The normal force is $F_{\rm N}=250 \ {\rm N}$ (red data points), $430 \ {\rm N}$ (green), and 
$250 \ {\rm N}$ (blue). For the first two cases, the experiments are performed with the 
same rubber blocks with an area of $A_0=7 \ {\rm cm^2}$, while the third experiment is performed 
with different rubber blocks (after run-in) with an area of $A_0=16 \ {\rm cm^2}$. 
The nominal contact pressures $F_{\rm N}/A_0$ are indicated in the figure. 
The average friction coefficient is $\mu \approx 0.95$, and the average wear rate is 
$\Delta m/F_{\rm N}L \approx 0.023 \ {\rm mg/Nm}$, this corresponds to an average wear volume $\Delta V/F_{\rm N}L \approx 0.019 \ {\rm mm^3/Nm}$.

\begin{table}[ht!]
\renewcommand{\arraystretch}{1.3}
\centering
\begin{tabular}{|l||c|c|c|}
\hline
\multirow{2}{*}{System} & \multicolumn{1}{c|}{$\sigma_0$} & \multirow{2}{*}{Concrete type} & \multicolumn{1}{c|}{Wear rate} \\
 & [MPa] & & [mg/Nm] \\
\hline
\hline
SBRc, dry & \multirow{4}{*}{0.36} & \multirow{2}{*}{Smooth} & $0.023,0.024$ \\
SBRc, in water & & & $0.038,0.036$ \\
SBRc, dry & & \multirow{2}{*}{Rough} & $0.038$ \\
SBRc, in water & & & $0.061$ \\
\hline
SBRc, dry & 0.61 & Smooth & $0.023$ \\
\hline
SBRc (3 blocks), dry & 0.16 & Smooth & $0.023$ \\
\hline
\hline
SBRs, dry & \multirow{4}{*}{0.36} & \multirow{2}{*}{Smooth} & $0.023,0.024$ \\
SBRs, in water & & & $0.038,0.058$ \\
SBRs, dry & & \multirow{2}{*}{Rough} & $0.027$ \\
SBRs, in water & & & $0.058$ \\
\hline
SBRs, dry & 0.61 & Smooth & $0.023$ \\
\hline
SBRs (3 blocks), dry & 0.16 & Smooth & $0.023$ \\
\hline
\end{tabular}
\caption{Experimental conditions and wear test results of SBR samples. $\sigma_0$ indicates the nominal contact pressure. Dry and in-water conditions are compared for each load case.}
\label{SBR}
\end{table}

These wear rates are higher than those observed in a previous study (see Ref.  \cite{ToBe}) 
for NRc, where $\Delta V/F_{\rm N}L = 0.0085 \ {\rm mm^3/Nm}$ 
was found initially, and $\Delta V/F_{\rm N}L = 0.0058 \ {\rm mm^3/Nm}$ a few months later
on a concrete surface from a different batch, which may explain the differences in wear rates.

Fig. \ref{NewPicWearParticles.1.eps} shows optical images of the wear 
particles collected after sliding rubber blocks on concrete surfaces.
The rubber particles are collected using adhesive film pressed against the rubber surface 
after specially designed sliding tests were conducted to collect wear particles, 
where the nominal contact pressure is $\sigma_0 = 0.056 \ {\rm MPa}$. 
Fig. \ref{NewPicWearParticles.1.eps} (a) and (b) illustrate that both the silica-filled 
and carbon-filled SBR compounds produced similar wear particle distributions.

Fig. \ref{1s.2wearrate.WATER.eps}(a) shows the friction coefficient $\mu$, 
and Fig. \ref{1s.2wearrate.WATER.eps}(b) shows the mass loss per unit normal force, $\Delta m/F_{\rm N}$, 
for silica- and carbon-filled SBR blocks as a function of sliding distance on concrete surfaces. 
The red symbols represent the dry state, and the green symbols represent the in-water state. 
The filled and open squares represent SBR with carbon and silica fillers, respectively. 
The same rubber blocks as in Fig. \ref{1s.2wearrate.all.loads.eps} were used. 
The nominal contact pressure is $0.36 \ {\rm MPa}$. 
The dashed lines in Fig. \ref{1s.2wearrate.WATER.eps}(b) correspond to wear rates 
$\Delta m/F_{\rm N}L \approx 0.038 \ {\rm mg/Nm}$ in water and $0.023 \ {\rm mg/Nm}$ in the dry state, corresponding to the wear volumes 
$\Delta V/F_{\rm N}L \approx 0.032 \ {\rm mm^3/Nm}$ and $0.019 \ {\rm mm^3/Nm}$, respectively
For the tests conducted in water, the wear debris is shown in 
Figs. \ref{ALLwearCarbon.eps}, \ref{ALLwearSilicone.eps}, and \ref{SBcPerticlesWater.ps}.

Fig. \ref{ALLwearCarbon.eps} shows two wear tracks for carbon-filled SBR blocks sliding on concrete surfaces in water.
Each test consisted of 10 sliding cycles, with each cycle comprising a $20 \ {\rm cm}$ forward and backward motion. 
Similar results for silica-filled SBR are presented in Fig. \ref{ALLwearSilicone.eps}(a). 
Fig. \ref{ALLwearSilicone.eps}(b) provides a magnified view of wear particles accumulated 
at the edge of one sliding track.

Wear particles formed during sliding on dry surfaces may agglomerate and form larger particles, a phenomenon that is less likely to occur during sliding in water. Fig. \ref{SBcPerticlesWater.ps} shows wear particles collected on filter paper from the carbon black-filled SBR compound sliding in water on the concrete surface. The size distribution of these particles is similar to that observed in the dry sliding configuration.

Fig. \ref{ZZ1distance.2mass.eps}(a) shows mass loss of rubber blocks as a function of sliding distance for SBR with carbon black filler (filled squares) and silica filler (open squares) on dry concrete surfaces. The red and green symbols correspond to the smooth and rough concrete surfaces, respectively. Two sliding cycles (each consisting of $17 \ {\rm cm}$ forward and backward motion) are performed for each case, and the sliding is repeated three times for each rubber/concrete combination (carbon-smooth, carbon-rough, silica-smooth, silica-rough).

Fig. \ref{ZZ1distance.2mass.eps}(b) shows the friction coefficients for the systems studied in Fig. \ref{ZZ1distance.2mass.eps}(a). The squares represent the friction coefficients for each sliding cycle ($2\times 3 = 6$ cycles for each rubber/concrete combination), and the circles indicate the average friction coefficient for each combination. The average values are 1.019, 1.052, 1.071, and 1.026 for the carbon-smooth, carbon-rough, silica-smooth, and silica-rough systems, respectively.

Fig. \ref{XX1distance.2wearmass.SB.wet.eps} shows similar results to Fig. \ref{ZZ1distance.2mass.eps}, but for sliding in water. Notably, the wear rates in water are significantly higher, by factors of $2-4$, compared to the dry state.

It is remarkable that the wear rates for SBR with both silica and carbon fillers in water are much higher than in the dry state. This result contrasts with findings from an earlier study on NR with carbon fillers, where the wear rate in water is much lower than in the dry state \cite{ToBe}. To explore this discrepancy further, we conducted additional tests on the same NR compound used in Ref. \cite{ToBe}.

\begin{figure}[h]
\includegraphics[width=0.47\textwidth,angle=0.0]{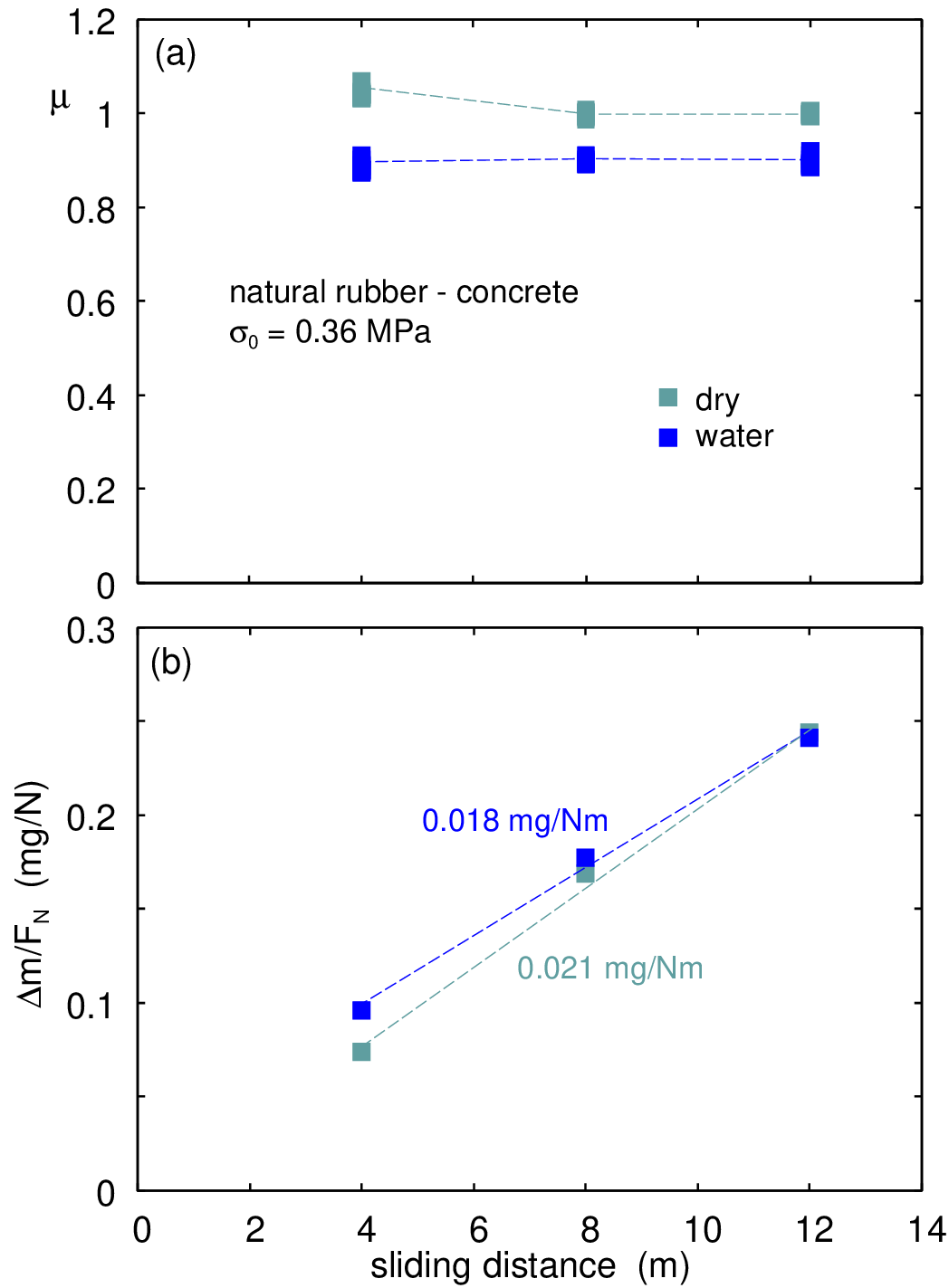}
\caption{\label{1distance.2wearmass.NR.dry.water.eps}
(a) The friction coefficient $\mu$ and (b) the mass loss divided by the normal force, $\Delta m/F_{\rm N}$,
for NR blocks as a function of the sliding distance on concrete surfaces at the nominal contact pressure of $\sigma_0 = 0.36 \ {\rm MPa}$. 
The gray and blue symbols represent the dry and in-water state. Dashed lines in (b) correspond to wear rates $\Delta m/F_{\rm N}L \approx 0.021  \ {\rm mg/Nm}$ in dry state and $ \approx 0.018  \ {\rm mg/Nm}$ in water, corresponds to 
$\Delta V/F_{\rm N}L \approx 0.018$ and $0.015 \ {\rm mm^3/Nm}$.
}
\end{figure}

\begin{figure}[h]
\includegraphics[width=0.47\textwidth,angle=0.0]{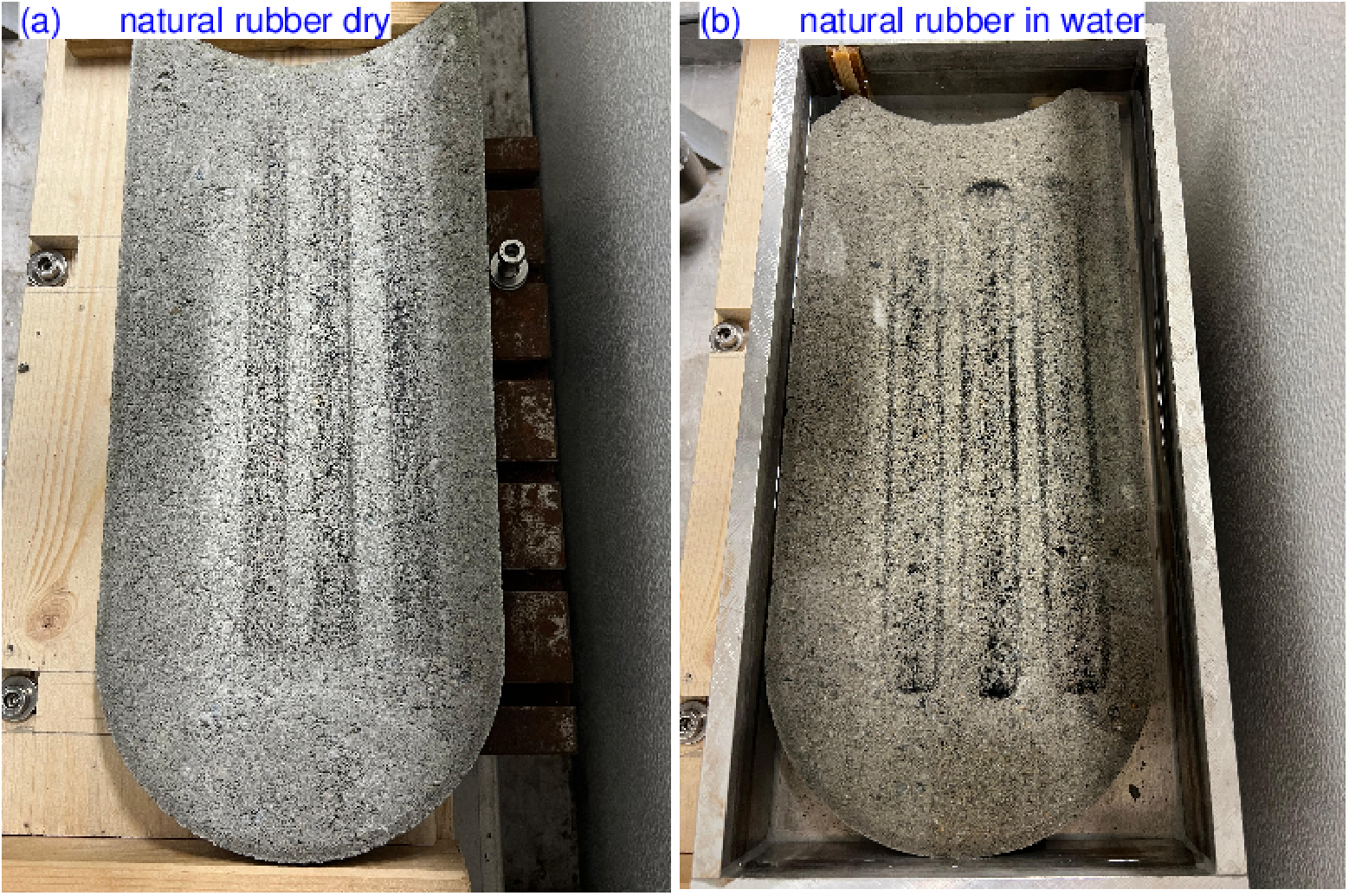}
\caption{\label{NRdrywetphotos.eps}
Wear tracks for the carbon-filled NR block sliding on a concrete surface in water.
The sliding motion consisted of 10 sliding cycles, each comprising a $20 \ {\rm cm}$ forward and backward motion.
}
\end{figure}

\begin{figure}[h]
\includegraphics[width=0.45\textwidth,angle=0.0]{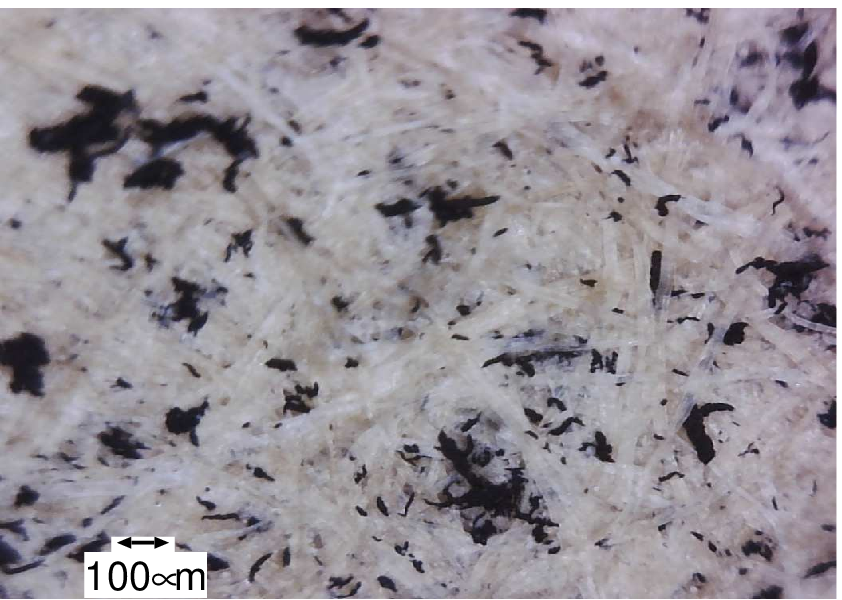}
\caption{\label{NRwetPHOTO.ps}
Wear particles from the experiment with NR in water on a  
concrete surface (Fig. \ref{1distance.2wearmass.NR.dry.water.eps}), collected on filter paper.
}
\end{figure}

\begin{figure}[h]
\includegraphics[width=0.47\textwidth,angle=0.0]{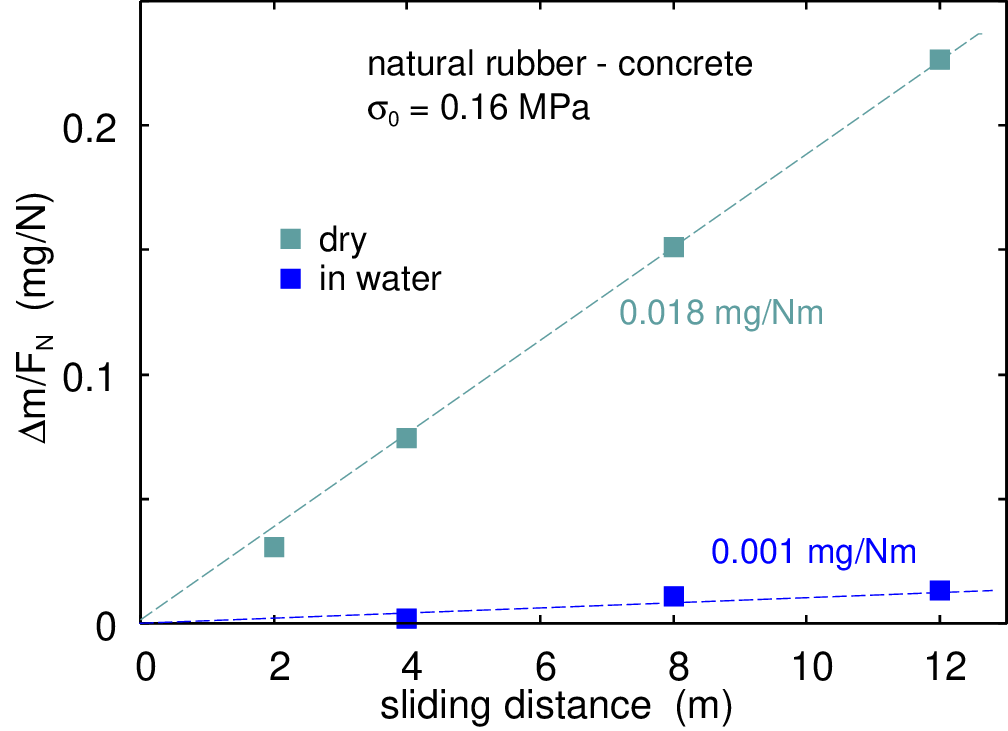}
\caption{\label{1distance.2mass.NR.dry.water.new.eps}
The mass loss divided by the normal force, $\Delta m/F_{\rm N}$,
for NR blocks as a function of the sliding distance on concrete surfaces at the nominal contact pressure of $\sigma_0 =  0.16 \ {\rm MPa}$. 
The gray and blue symbols correspond to the dry and in-water state. The dashed lines correspond to wear rates 
$\Delta m/F_{\rm N}L \approx 0.018$ and $ 0.001  \ {\rm mg/Nm}$ in the dry and in-water state. The average friction coefficients for dry and in-water scenarios are $\mu = 1.029$ and $0.919$, respectively.
	}
\end{figure}

\begin{figure}[h]
\includegraphics[width=0.47\textwidth,angle=0.0]{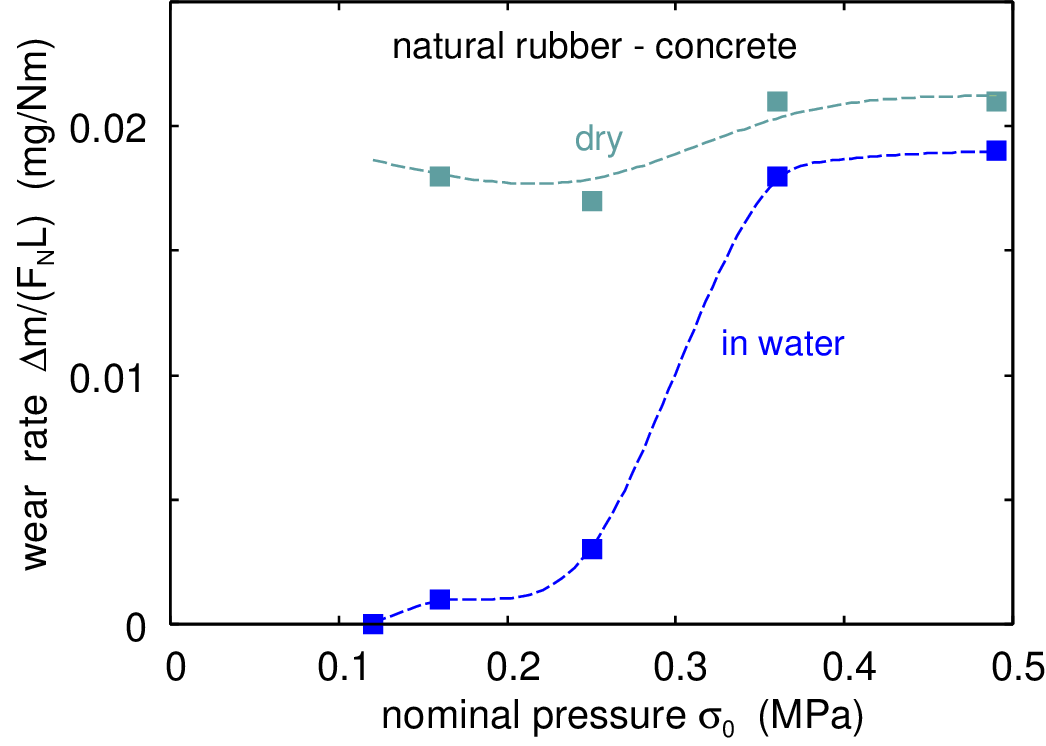}
\caption{\label{1pressure.2wearate.NRcarbon.dry.water.eps}
The mass loss normalized by the normal force, $\Delta m/F_{\rm N}$,
for NR blocks as a function of the nominal contact pressure. 
The gray and blue symbols correspond to the dry and in-water state.}
\end{figure}

\vskip 0.2cm
{\bf NR on concrete, dry and in water}

Experimental conditions and results for NRc are summarized in Table. \ref{NR} and shown in Fig. \ref{1distance.2wearmass.NR.dry.water.eps}. Two samples with different surface areas, $A_0 = 16 \ {\rm cm^2}$ (three blocks) and $7 \ {\rm cm^2}$ (single block), are tested under three different loads, $F_{\rm N} = 250$, $340$, and $400 \ {\rm N}$, to achieve nominal contact pressures of $\sigma_0 = 0.16$, $0.25$, $0.36$, and $0.49 \ {\rm MPa}$. 

Fig. \ref{1distance.2wearmass.NR.dry.water.eps} (a) shows the friction coefficient $\mu$ and (b) the mass loss normalized by the normal force, $\Delta m/F_{\rm N}$, for NR blocks as a function of the sliding distance on concrete surfaces. The gray symbols represent the dry state, while the blue symbols correspond to in-water tests. The nominal contact pressure is $0.36 \ {\rm MPa}$. Wear debris collected from the in-water test is illustrated in Fig. \ref{NRdrywetphotos.eps} and Fig. \ref{NRwetPHOTO.ps}.

\begin{table}[ht!]
\renewcommand{\arraystretch}{1.3}
\centering
\begin{tabular}{|l||c|c|c|}
\hline
\multirow{2}{*}{System} & \multicolumn{1}{c|}{$\sigma_0$} & \multirow{2}{*}{Concrete type} & \multicolumn{1}{c|}{Wear rate} \\
 & [MPa] & & [mg/Nm] \\
\hline
\hline
NRc, dry & \multirow{2}{*}{0.36} & \multirow{8}{*}{smooth} & $0.021$ \\
NRc, in water & & & $0.018$ \\
\cline{1-2} \cline{4-4}
NRc, dry & \multirow{2}{*}{0.49} & & $0.021$ \\
NRc, in water & & & $0.019$ \\
\cline{1-2} \cline{4-4}
NRc (3 blocks), dry & \multirow{2}{*}{0.16} & & $0.018$ \\
NRc (3 blocks), in water & & & $0.0012$ \\
\cline{1-2} \cline{4-4}
NRc (3 blocks), dry & \multirow{2}{*}{0.25} & & $0.017$ \\
NRc (3 blocks), in water & & & $0.0030$ \\
\hline
\end{tabular}
\caption{Experimental conditions and wear test results of NRc samples on smooth concrete. 
$\sigma_0$ represents the nominal contact pressure. Dry and in-water conditions are compared for each load case.}
\label{NR}
\end{table}

Fig. \ref{1distance.2mass.NR.dry.water.new.eps} presents similar results to Fig. \ref{1distance.2wearmass.NR.dry.water.eps}(b), but for a lower nominal contact pressure of $\sigma_0 = 0.16 \ {\rm MPa}$. The dashed lines indicate wear rates of $\Delta m/F_{\rm N}L \approx 0.018 \ {\rm mg/Nm}$ in the dry state and $0.001 \ {\rm mg/Nm}$ in water. Notably, the wear rate in the dry state remains consistent with that observed at $\sigma_0 = 0.36 \ {\rm MPa}$, which is consistent with the results of Ref. \cite{ToBe}, where no dependency of the wear rate on nominal contact pressure in the dry state is reported. However, the wear rate in water is significantly smaller at $\sigma_0 = 0.16 \ {\rm MPa}$ compared to $0.36 \ {\rm MPa}$.

Fig. \ref{1pressure.2wearate.NRcarbon.dry.water.eps} summarizes the observed wear rates in both the dry state and in water as a function of the nominal contact pressure $\sigma_0$. In water, the wear rate decreases sharply and approaches zero as the nominal contact pressure decreases to below $\sim 0.3 \ {\rm MPa}$.

To summarize, the study reveals that the wear rate normalized by the applied normal force in the dry state is independent of the nominal contact pressure (or normal force) within the tested pressure range ($0.16-0.61 \ {\rm MPa}$ for SBR and $0.16-0.49 \ {\rm MPa}$ for NR compounds). This finding aligns with theoretical predictions developed in Ref. \cite{ToBe}. Conversely, in water, the wear rate for SBR is $2-4$ times higher than in the dry state, while for NR, it is significantly lower and vanishes for $\sigma_0 < 0.12 \ {\rm MPa}$.

\begin{figure}[ht!]
\includegraphics[width=0.25\textwidth,angle=0.0]{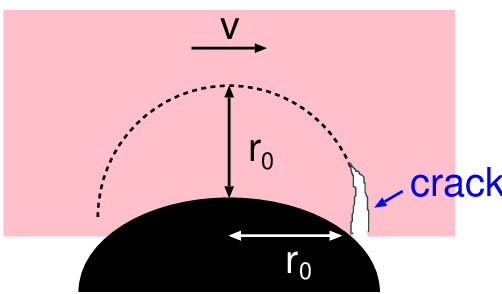}
\caption{\label{WearParticlePMMA.eps}
Schematic of a rubber block sliding in contact with a hard countersurface. 
The sliding speed $v$ and the radius of the contact region $r_0$ are indicated. 
The deformation field extends both laterally and vertically into the polymer over a comparable distance.
}
\end{figure}

\vskip 0.3cm
{\bf 4 Review of theory of sliding wear}

Sliding wear depends on the size of the contact regions and the stresses acting within these regions \cite{Mu1}. In Ref. \cite{ToBe}, we developed a theory of sliding wear. Here, we review the main results using a slightly different approach.

Cracks at the surface of a solid can be induced by both normal and tangential stresses acting on the surface. However, if the friction force is comparable to the normal force, particle removal is primarily driven by tangential stress. This is because the normal force predominantly induces compressive deformations, which are not effective in driving crack opening. 

Even in the absence of friction, tensile stresses can arise outside the contact regions, potentially generating circular or radial opening cracks if the deformation field is sufficiently large. This mechanism can significantly influence the wear rate, particularly when the frictional shear stress is small. This topic is further explored in Appendix C, where it is also related to studies of crack formation during normal indentation of brittle solids.

Let $\tau = \tau (\zeta_r)$ denote the effective shear stress acting in an asperity contact region with a radius $r_0$. 
The magnification $\zeta_r$ is determined by the radius of the contact region, $q_r = \pi /r_0$, and $\zeta_r = q_r /q_0$, 
where $q_0=2\pi/L$, with $L$ the linear size of the system. 
The elastic energy stored in the deformed asperity contact is (see Fig. \ref{WearParticlePMMA.eps}):
$$U_{\rm el} \approx {\tau^2  \over  E^*} r_0^3,$$
where the effective modulus is $E^* = E/(1-\nu^2)$ (assuming the substrate is rigid). 

More accurately, if we assume the shear stress acts uniformly within a circular region of radius $r_0$, 
the center of this circular region will displace a distance $u$ given by $k u = F$, where $F=\tau \pi r_0^2$ is the force, 
and $k \approx (\pi/2) E^* r_0$ is the spring constant. 
This gives the elastic energy:
$$U_{\rm el} = {1\over 2} k u^2 = {F^2 \over 2 k} = 
{(\pi r_0^2 \tau)^2\over \pi E^* r_0} = \pi {\tau^2 \over E^*} r_0^3.\eqno(1)$$

For the shear stress to remove a particle of linear size $r_0$, the stored elastic energy must be larger than the fracture (crack) energy which is of the order:
$$U_{\rm cr} \approx \gamma 2 \pi r_0^2, \eqno(2)$$
where $\gamma$ is the energy per unit surface area required to break bonds at the crack tip. 
If $U_{\rm el} > U_{\rm cr}$, the elastic energy is sufficient to propagate a crack and remove a particle \cite{Rabi1,Rabi2,Moli1}. 
Thus, for a particle to be removed, we must have $\tau > \tau_{\rm c}$:
$$\tau_{\rm c} = \beta \left ({2 E^* \gamma \over r_0 }\right )^{1/2}, \eqno(3)$$
where $\beta$ is a number of order unity, accounting for the fact that wear particles are not generally hemispherical, as assumed above.

In the following, we treat the rubber surface as smooth and assume the counter surface is rough. 
A substrate asperity, where the shear stress is sufficient to remove a particle of size $r_0$, is referred to as a {\it wear-asperity}, 
and the corresponding contact region as the {\it wear-contact region}.

Assuming that the effective shear stress $\tau$ is proportional to the normal stress $\sigma$, $\tau = \mu \sigma$, 
particles will only be removed if the contact stress $\sigma > \sigma_{\rm c} (\zeta)$, where:
$$\sigma_{\rm c} = {\beta \over \mu} \left ({2 E^* \gamma \over r_0 }\right )^{1/2}. \eqno(4)$$

For randomly rough surfaces under elastic contact conditions, the probability distribution of contact stress is given by:
$$P(\sigma,\zeta) = {1\over (4\pi G)^{1/2}} \left (e^{-(\sigma-\sigma_0)^2/4G}- e^{-(\sigma+\sigma_0)^2/4G} \right ), \eqno(5)$$
where $\sigma_0$ is the nominal (applied) pressure, and:
$$G ={\pi \over 4} (E^*)^2 \int_{q_0}^{\zeta q_0} dq \ q^3 C(q), \eqno(6)$$
with $C(q)$ being the surface roughness power spectrum.

When the interface is studied at the magnification $\zeta$, 
the area $A = A_{\rm wear}(\zeta)$, where the shear stress is sufficient to remove particles, is given by:
$$
{A_{\rm wear} (\zeta) \over A_0} = \int_{\sigma_{\rm c} (\zeta)}^\infty d\sigma \ P(\sigma,\zeta). \eqno(7)
$$

At magnification $\zeta$, the smallest wear particles observed have a size $r_0 \approx \pi /q_r$, where $q_r = \zeta q_0$. 
Thus, at this magnification, the pixel size is $r_0 = \pi /\zeta q_0$, and the smallest removed particle observable is determined by the pixel size.

The crack energy $\gamma$ depends on the bond-breaking speed and spans a range of values, 
$\gamma_0 < \gamma < \gamma_1$. The faster the crack propagates, the larger $\gamma$ becomes. 
The smallest stored elastic energy, $U_{\rm el} = U_{\rm el 0}$, required to remove a particle is given by 
$U_{\rm el 0} \approx \gamma_0 2\pi r_0^2$. In this case, the crack moves extremely slowly, and the incremental displacement $\Delta x$ 
during the interaction between the moving {\it wear-asperity} and the crack is very small, 
requiring many $\sim r_0/\Delta x$ contacts to remove the particle. If the interaction with a {\it wear-asperity} results in $U_{\rm el} \gg U_{\rm el 0}$, 
the crack moves much faster (larger $\Delta x$), and far fewer contacts are needed to remove a particle.

During sliding, the crack interacts with many {\it wear-asperities} of varying sizes, 
experiencing a wide range of crack-tip displacements $\Delta x$ before the particle is ultimately removed.

The probability that the stress at an arbitrary point on the polymer surface lies between $\sigma$ and $\sigma + d\sigma$, 
when the interface is studied at magnification $\zeta$, is given by $P(\sigma,\zeta) d\sigma$. 
If $\sigma > \sigma_{\rm c}$, the local stress results in $U_{\rm el} > U_{\rm el 0}$, so a particle can potentially be removed. 
However, during the interaction time, the crack moves only the distance $\Delta x (\gamma)$, where the relevant $\gamma$ 
is assumed to satisfy $U_{\rm el} = \gamma 2\pi r_0^2$. Hence, the number of contacts needed to remove the particles is $N(\gamma) = r_0/\Delta x$. 
After the run-in, the probability that a particle is removed from regions where the stress is in the range $\sigma$ to $\sigma + d\sigma$ 
is $P(\sigma,\zeta) d\sigma /N(\gamma)$. The total probability becomes:
$$
P^*=\int_{\sigma_{\rm c}}^\infty d\sigma {P(\sigma, \zeta) \over 1+ r_0(\zeta )/\Delta x(\sigma, \zeta)}. \eqno(8)
$$
where we add 1 in the denominator to ensure the correct limit as $\Delta x/r_0 \rightarrow \infty$.

In   (8), the cut-off stress $\sigma_{\rm c}$ is determined by $U_{\rm el} = U_{\rm el 0}$. 
There are $N^* = A_0/\pi r_0^2$ pixels on the surface. Sliding a distance $L = 2r_0$ 
removes $N^* P^*$ particles, corresponding to a volume $V = (2 \pi r_0^3/3) N^* P^*$. 
Thus, we get $V/L = (\pi r_0^2/3) N^* P^*$ or:
$$
{V \over LA_0} = {P^* \over 3}.
$$
Using   (8), this becomes:
$$
{V\over LA_0} = {1\over 3} \int_{\sigma_{\rm c}}^\infty d\sigma {P(\sigma, \zeta) \over 1+ r_0(\zeta )/\Delta x(\sigma, \zeta)}. \eqno(9)
$$
This expression matches (13) in Ref.  \cite{ToBe}, except that the factor of $1/2$ in Ref.  \cite{ToBe} is replaced by $1/3$ in   (9) 
due to a slightly different description of the particle removal process.

The number of contacts needed to remove a particle, $N_{\rm cont} \approx r_0/\Delta x$, 
depends on the crack energy $\gamma$ and could be very large ($10^2$ or more) 
if the macroscopic relation between the tearing energy $\gamma$ and $\Delta x$ also holds at the wear particle scale.

The theory above provides the wear volume assuming that particles of a given size (radius $r_0$) are generated.
These correspond to the smallest particles observable at the magnification $\zeta = q_r/q_0 = \pi / q_0 r_0$.
To calculate the total wear volume, we must sum the volumes of wear particles across all length scales, 
which can be observed as the magnification increases. To avoid double-counting particles of similar size, 
we increment the magnification in steps of $2$, defining $\zeta = 2^n = \zeta_n$, 
where $n = 0, 1, \dots, n_1$, and $2^{n_1} q_0 = q_1$. 
The interval from $\zeta = 2^n$ to $2^{n+1}$ is referred to as a 2-interval.

In Ref.  \cite{ToBe}, it was shown that:
$$
{V\over A_0 L} = {1\over 3 {\rm ln}2 } \int_{q_0}^{q_1} dq  
\int_{\sigma_{\rm c} (\zeta)}^\infty d\sigma \ 
{P(\sigma,\zeta) \over q+\pi/\Delta x(\sigma,\zeta)}, \eqno(10)
$$
where $\zeta = q/q_0$. By substituting $q = q_0 e^\xi$ so that $dq = q d\xi$, this can be rewritten as:
$$
{V\over A_0 L} = {1\over 3 {\rm ln}2 } \int_{0}^{\xi_1} d\xi  
\int_{\sigma_{\rm c} (\zeta)}^\infty d\sigma \ 
{P(\sigma,\zeta) \over 1+r_0(\zeta)/\Delta x(\sigma,\zeta)}. \eqno(11)
$$
where $\xi_1 = {\rm ln}(q_1/q_0)$.

The number of particles with radius $r_0$ between $(\pi / q_0) 2^{-n-1/2}$ and $(\pi / q_0) 2^{-n+1/2}$ is given by:
$$
{N_n \over A_0 L} \approx {1 \over 3 \pi r_0^3 (\zeta_n) } 
\int_{\sigma_{\rm c} (\zeta_n)}^\infty d\sigma \ 
{P(\sigma,\zeta_n) \over 1+r_0(\zeta_n)/\Delta x(\sigma,\zeta_n)}. \eqno(12)
$$

If $r_0/\Delta x$ is large, a long run-in distance may be required before the wear process reaches a steady state.
This is particularly true when the nominal contact pressure is small, leading to large distances between the wear asperity 
contact regions. However, in cases where the contact regions within the macroasperity contacts are densely distributed 
and independent of the nominal contact pressure, sufficient wear asperity contact regions within the macroasperity contacts 
may exist to achieve the $N_{\rm cont}$ needed for wear particle formation even over a short sliding distance.

\begin{figure}
\includegraphics[width=0.47\textwidth,angle=0.0]{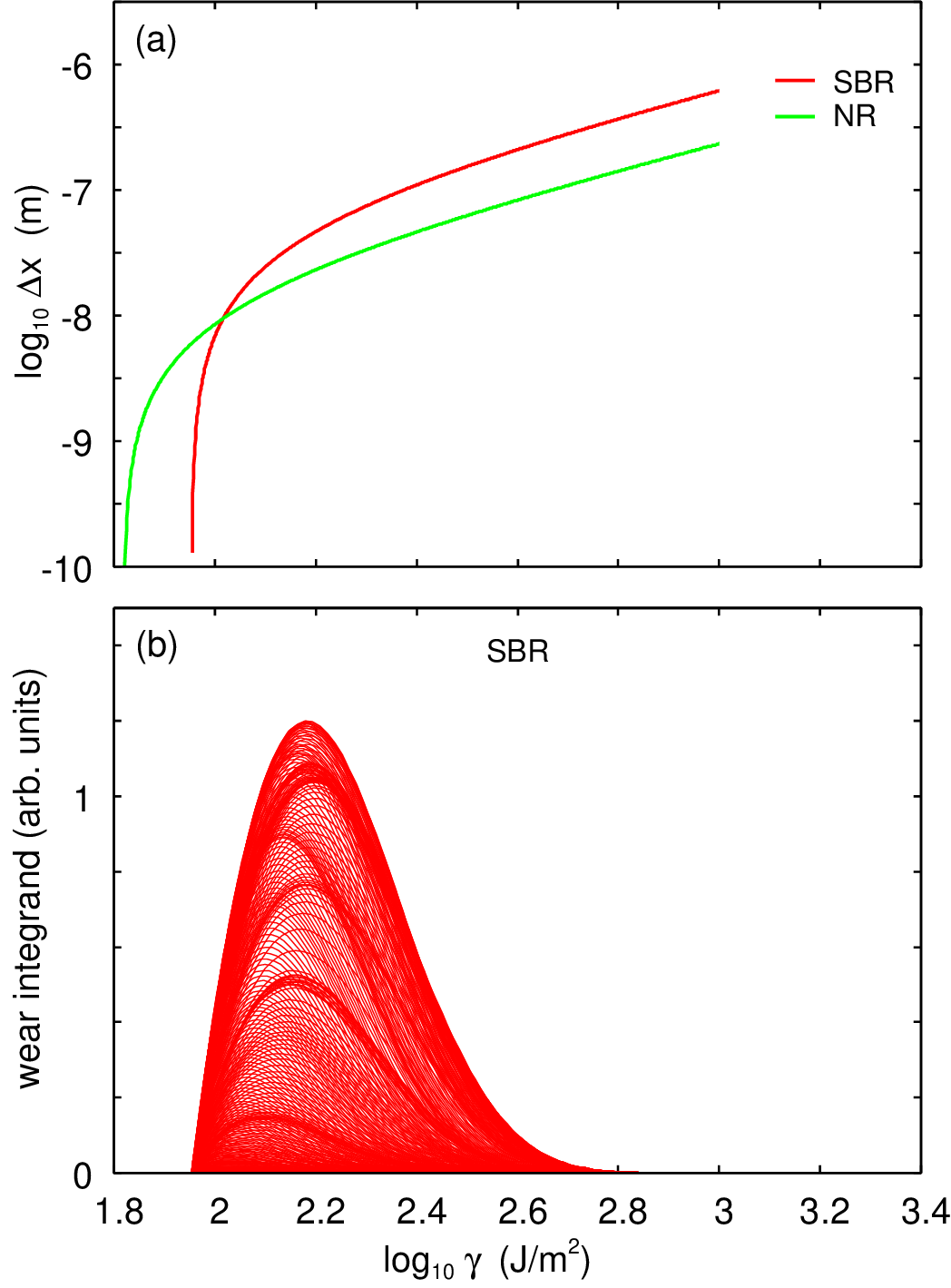}
\caption{\label{1logGamma.2WearIntegrand.SBR.eps}
(a) The relationship between the crack tip displacement per oscillation and the tearing (or crack) energy 
$\gamma$ for SBR (red line) and NR (green line) used in the calculations. 
(b) The integrand in   (9) as a function of $\gamma$ for all magnifications
(or particle radii $r_0$) for SBR sliding on concrete.
[Note: The integration variable in   (9) is the pressure (or stress), but each pressure corresponds to a 
tearing energy as determined by   (4).] The red area represents the superposition of many curves for the different
magnifications or particle radii.
We have used $E=20 \ {\rm MPa}$, $\nu = 0.5$, and the relation between the crack-tip displacement $\Delta x (\gamma)$
and the tearing energy $\gamma$ is shown in (a) (red line).
The nominal contact area $A_0 = 16 \ {\rm cm^{2}}$ and the nominal contact pressure $\sigma_0 = 0.36 \ {\rm MPa}$
as in the experiment described in Sec. 3. The friction coefficient $\mu = 0.95$ is also used.
}
\end{figure}

\begin{figure}
\includegraphics[width=0.47\textwidth,angle=0.0]{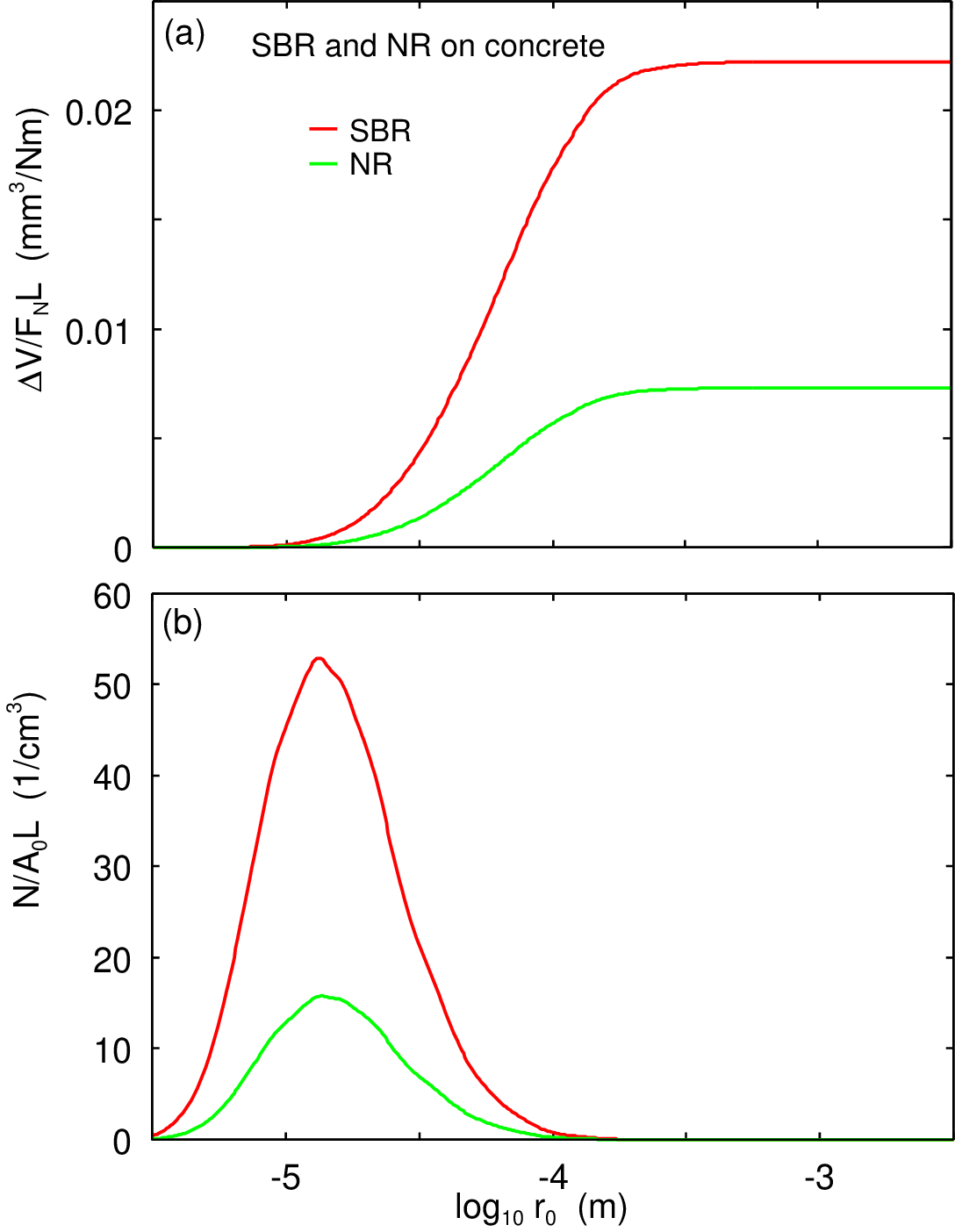}
\caption{\label{FINAL1logr0.2WearVolume.overNm.smooth.SBR.NR.eps}
(a) The cumulative wear rate and
(b) the distribution of the number of generated particles,
as a function of the logarithm of the particle radius for SBR (red line) and NR (green line) sliding on concrete.
For SBR, we used $E=20 \ {\rm MPa}$, $\nu = 0.5$, and the relationship between the crack-tip displacement $\Delta x (\gamma)$
and the tearing energy $\gamma$ shown in Fig. \ref{1logGamma.2WearIntegrand.SBR.eps}(a) (red line).
For NR, we used $E=14 \ {\rm MPa}$, $\nu = 0.5$, and the relationship between the crack-tip displacement $\Delta x (\gamma)$
and the tearing energy $\gamma$ shown in Fig. \ref{1logGamma.2WearIntegrand.SBR.eps}(a) (green line).
The calculated wear rates for the SBR and NR compounds are $0.0222$ and $0.0073 \ {\rm mm^3/Nm}$, respectively.
The nominal contact pressure $\sigma_0 = 0.36 \ {\rm MPa}$.
}
\end{figure}

\begin{figure}
\includegraphics[width=0.47\textwidth,angle=0.0]{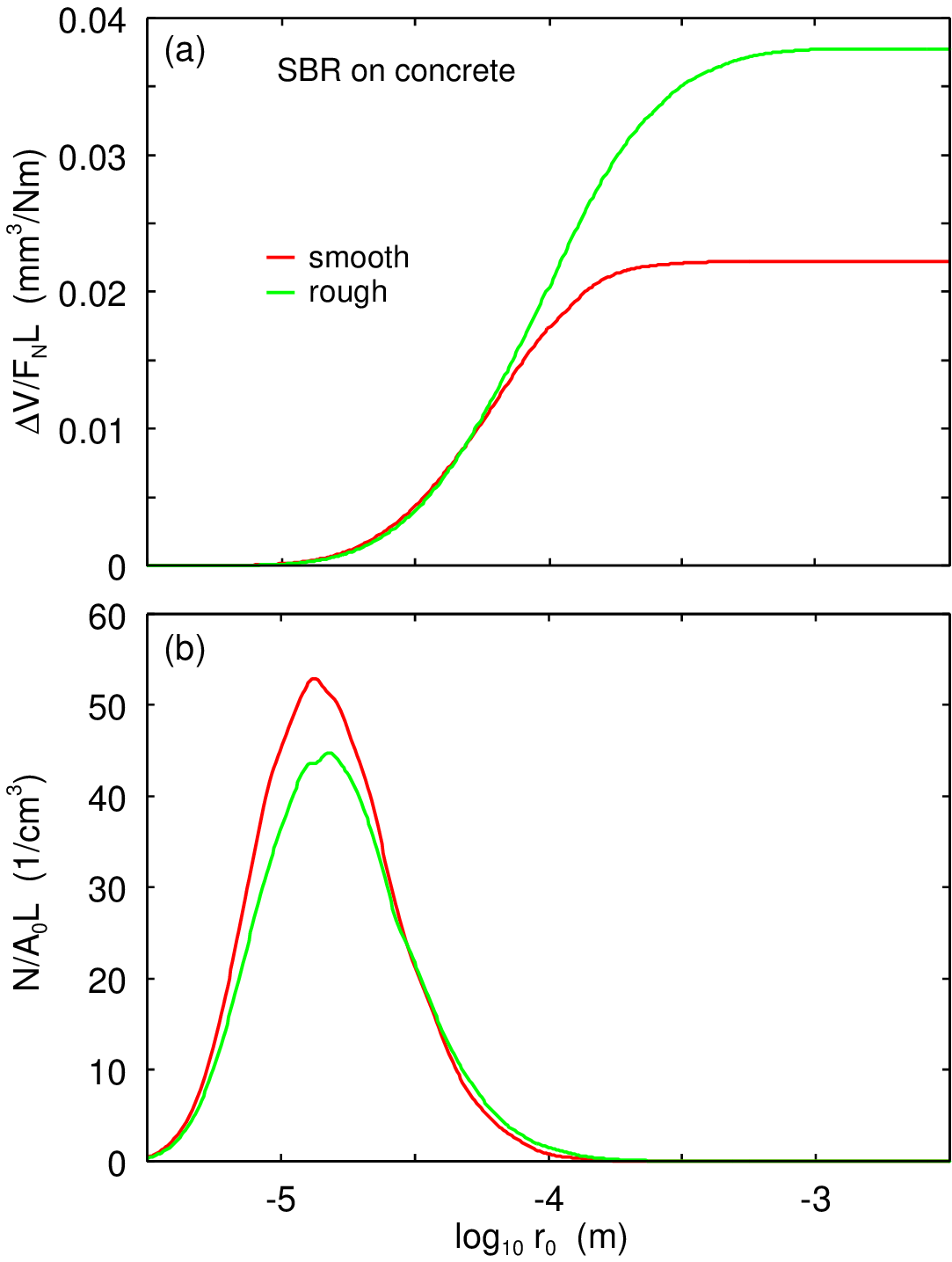}
\caption{\label{FINAL1logr0.2WearVolume.overNm.smooth.rough.eps}
(a) The cumulative wear volume and (b) the number of generated particles,
as a function of the logarithm of the particle radius for the SBR compound sliding on ``smooth'' and ``rough'' concrete surfaces.
The calculations are performed using $E=20 \ {\rm MPa}$, $\nu = 0.5$, and the relationship between the crack-tip displacement $\Delta x (\gamma)$
and the tearing energy $\gamma$ shown in Fig. \ref{1logGamma.2WearIntegrand.SBR.eps}(a) (red line).
The calculated wear rates for the ``smooth'' and ``rough'' surfaces are $0.0222$ and $0.0378 \ {\rm mm^3/Nm}$, respectively.
The nominal contact pressure $\sigma_0 = 0.36 \ {\rm MPa}$.
}
\end{figure}

\vskip 0.3cm
{\bf 5 Comparing Theory with Experiments}

Here we compare the theoretical predictions for the wear rate with the experimental results for 
rubber sliding on dry concrete surfaces.
The relationship between $\Delta x$ and the tearing energy $\gamma$ has been measured for SBR with different amounts
of carbon black and silica filler (see, e.g., Fig. 7 in Ref.  \cite{Tear,RGK}). Here, we have used the general
relationship between the crack-tip displacement $\Delta x$ and the tearing energy $\gamma$ observed in experiments,
which, for the relatively small $\gamma$ of interest here, can be approximated as
$$\Delta x = 0 \ \ \ \ \ \ {\rm for} \ \gamma < \gamma_0\eqno(13)$$
$$\Delta x = a\times (\gamma-\gamma_0), \ \ \ \ \ \ {\rm for} \ \gamma_0 < \gamma < \gamma_1\eqno(14)$$
We use the following parameters for SBR:
$$E= 20 \ {\rm MPa}, \ \ \ \gamma_0 = 90 \ {\rm J/m^2}, \ \ \ a = 5.6 \times 10^{-10} \ {\rm m^3 /J}$$ 
and for NR:
$$E= 14 \ {\rm MPa}, \ \ \ \gamma_0 = 66 \ {\rm J/m^2}, \ \ \ a = 2.0 \times 10^{-10} \ {\rm m^3 /J}$$ 
For $\gamma$ larger than $\gamma_1$, which is typically on the order of a few hundred ${\rm J/m^2}$, $\Delta x$ depends on
$\gamma$ as a power-law relation, and as $\gamma$ approaches $\gamma_2 \approx 10^5 \ {\rm J/m^2}$, the crack displacement $\Delta x \rightarrow \infty$.
In the present application, $\gamma < 500 \ {\rm J/m^2}$ (see Fig. \ref{1logGamma.2WearIntegrand.SBR.eps}), 
and we will use (13) and (14) for all $\gamma$.

Using these parameters, the friction coefficient $\mu = 0.95$, and the modulus $E=20 \ {\rm MPa}$,
we calculate for the concrete the cumulative wear rates 
shown in Fig. \ref{FINAL1logr0.2WearVolume.overNm.smooth.SBR.NR.eps}(a). The total wear rates
are $\Delta V/F_{\rm N} L \approx 0.0222$ and $0.0073 \ {\rm mm^3/Nm}$
for the SBR and NR compounds, respectively, which agree well with the experimental observations in the dry state where 
$\Delta V/F_{\rm N} L \approx 0.023$ and $\approx 0.006-0.008 \ {\rm mm^3/Nm}$ for the SBR and NR, respectively. 

We have also calculated the wear rate for the SBR compounds on the ``smooth'' and ``rough'' concrete surfaces.
Fig. \ref{FINAL1logr0.2WearVolume.overNm.smooth.rough.eps}
shows the cumulative wear volume (a) and the number of generated particles (b),
as a function of the logarithm of the particle radius. The ratio between the wear rates on the rough
and smooth surfaces is $0.0378/0.0222 \approx 1.7$. In this calculation, we have assumed the same
friction coefficient $\mu = 0.95$ on both surfaces.

\begin{table}[hbtp]
\renewcommand{\arraystretch}{1.8}
\centering
      \noindent\makebox[\linewidth][c]{%
      \begin{tabular}{@{}|l||c|c|c|c|c|@{}}
\hline
	      System & Dry (mg/Nm)  & In water (mg/Nm) \\
\hline
\hline
SBRc, smooth 
& \makecell{$0.023^{(1,2)}; \ 0.023$\\
	     $0.024$} 
	      & $0.038; \ 0.036$ \\
\hline
SBRc, rough & $0.038$ & $0.061$ \\
\hline
SBRs, smooth 
& \makecell{$0.023^{(1,2)}; \ 0.023$\\
	      $0.024$}
	      & $0.038; \ 0.058$ \\
\hline
SBRs, rough & $0.027$ & $0.058$ \\
\hline
SBRc [a], smooth & $0.014^{(3)}$ & $0.050^{(3)}$ \\
\hline
\hline
NRc, smooth 
& \makecell{$0.018^{(1)}; \ 0.017^{(4)}$\\
             $0.021;\ 0.021^{(5)}$} 
& \makecell{$0.0012^{(1)}; \ 0.003^{(4)}$\\
             $0.018;\ 0.019^{(5)}$} \\
\hline
NRc [b], smooth & $0.01^{(6)}; 0.007^{(6)}$ & $0.0^{(6)}$ \\
\hline
\end{tabular}
}
\caption{
Summary of wear rates for SBRs, SBRc, and NRc.  
[a] refers to an unpublished SBRc compound with a much higher glass transition temperature  
($-7^\circ {\rm C}$ compared to $-50^\circ {\rm C}$ for the SBR compound used in this study). 
[b] corresponds to the study reported in Ref.  \cite{ToBe} using the same NRc compound as in this study.  
For all systems, the friction coefficient is close to 1. The nominal contact pressure $\sigma_0 = 0.36 \ {\rm MPa}$, 
except for $(1) = 0.16 \ {\rm MPa}$, $(2) = 0.61 \ {\rm MPa}$, $(3) = 0.66 \ {\rm MPa}$, 
$(4) = 0.25 \ {\rm MPa}$, $(5) = 0.49 \ {\rm MPa}$ and $(6) = 0.12 \ {\rm MPa}$.}
\label{TABLE1}
\end{table}

\begin{table}[hbt]
\renewcommand{\arraystretch}{1.3}
\centering
      \begin{tabular}{@{}|l||c|c|c|c|c|@{}}
\hline
	      System & Ratio in-water/dry \\
\hline
\hline
SBRc, smooth & $1.65; \ 1.50$ \\
\hline
SBRc, rough & $1.60$ \\
\hline
SBRs, smooth & $1.65; \ 2.52$ \\
\hline
SBRs, rough & $2.15$ \\
\hline
SBRc [a], smooth & $3.57^{(3)}$ \\
\hline
\hline
NRc, smooth & $0.067^{(1)}; \ 0.18^{(4)}; \ 0.86; \ 0.91^{(5)}$ \\
\hline
NRc [b], smooth & $0.0^{(6)}$ \\
\hline
\end{tabular}
\caption{
The ratio of the wear rates for dry and in-water concrete for the systems 
presented in Table \ref{TABLE1}. The nominal contact pressure $\sigma_0 = 0.36 \ {\rm MPa}$, 
except for $(1) = 0.16 \ {\rm MPa}$, $(2) = 0.61 \ {\rm MPa}$, $(3) = 0.66 \ {\rm MPa}$, 
$(4) = 0.25 \ {\rm MPa}$, $(5) = 0.49 \ {\rm MPa}$ and $(6) = 0.12 \ {\rm MPa}$.}
\label{TABLE2}
\end{table}

\begin{table}[hbt]
\renewcommand{\arraystretch}{1.3}
\centering
      \begin{tabular}{@{}|l||c|c|c|c|c|@{}}
\hline
	      System & Ratio rough/smooth \\
\hline
\hline
SBRc, dry & $1.58$ \\
\hline
SBRc, in water & $1.69$ \\
\hline
SBRs, dry & $1.13$ \\
\hline
SBRs, in water & $1.00$ \\
\hline
\end{tabular}
\caption{\label{TABLE3}
The ratio of the wear rates for rough and smooth concrete for the systems 
presented in Table \ref{TABLE1}. The nominal contact pressure $\sigma_0 = 0.36 \ {\rm MPa}$.}
\end{table}

\vskip 0.3cm
{\bf 6 Discussion}

In two earlier studies on NR and SBR compounds, we measured the wear rate in both dry and in-water conditions. For the NR compound with carbon black filler, Ref. \cite{ToBe} reported no wear when sliding $25 \ {\rm m}$ in water on a concrete surface at a nominal contact pressure of $0.12 \ {\rm MPa}$ and sliding speed of $1 \ {\rm mm/s}$. 
However, under the same conditions in the dry state, a wear rate of $0.007 \ {\rm mg/Nm}$ was observed.

In the present study, using the same NR compound at a sliding speed of $3 \ {\rm mm/s}$ and nominal contact pressure of $0.36 \ {\rm MPa}$, which is approximately three times higher than in the earlier study, the wear test shown in Fig. \ref{1distance.2wearmass.NR.dry.water.eps} yielded a wear rate in water of $\approx 0.021 \ {\rm mg/Nm}$, comparable to that in the dry state ($\approx 0.018 \ {\rm mg/Nm}$). Subsequently, tests at additional nominal contact pressures, including two lower pressures of $0.16$ and $0.25 \ {\rm MPa}$, and a higher pressure of $0.49 \ {\rm MPa}$, revealed that the wear rates in the dry state
are nearly independent of the contact pressure, in the studied pressure range, 
ranging from $\approx 0.018$ to $0.021 \ {\rm mg/Nm}$. However, the wear rates in water exhibited significant variation, ranging from $\approx 0.001$ to $0.019 \ {\rm mg/Nm}$.

For the SBR compound, an unpublished study involved sliding a small compact rubber wheel made from an SBR compound ($T_{\rm g} \approx -7^\circ {\rm C}$, different from the one used in this study) on a concrete surface. This study found a wear rate of $0.012 \ {\rm mg/Nm}$ in the dry state and a fourfold increase, approximately $0.05 \ {\rm mg/Nm}$ in water. In this experiment, the average contact pressure was very high, about $0.66 \ {\rm MPa}$.

For the two SBR compounds used in the present study ($T_{\rm g} \approx -50^\circ {\rm C}$), we observed a wear rate in water that is approximately 1.7 times higher than in the dry state: about $\approx 0.038  \ {\rm mg/Nm}$ in water compared to $\approx 0.023  \ {\rm mg/Nm}$ in the dry state. This difference in wear rates between the two SBR compounds in water may reflect the different chemical compositions of the rubbers, as indicated by their significantly different glass transition temperatures. Additionally, the nominal contact pressures differed ($0.36 \ {\rm MPa}$ in the present study and $0.66 \ {\rm MPa}$ in the earlier study).

Table \ref{TABLE1} summarizes all the measured wear rates for SBR and NR sliding on concrete surfaces.  
Tables \ref{TABLE2} and \ref{TABLE3} provide the ratios of wear rates for dry and in-water concrete 
and for rubber sliding on rough and smooth concrete surfaces.
For the two studied SBR compounds, we observed significantly higher wear in water than in the dry state, with enhancement factors of $1.5-2.5$ for the low-glass-transition SBR compound 
($T_{\rm g} \approx -50^\circ {\rm C}$) and approximately $4$ for the higher glass-transition compound ($T_{\rm g} \approx -7^\circ {\rm C}$).

For the NR compound, almost no wear is observed in water at low nominal contact pressures ($\sigma_0 \approx 0.12$, $0.16$, and $ 0.25 \ {\rm MPa}$), while at $\sigma_0 \approx 0.36$ and $0.49 \ {\rm MPa}$, the wear rate is similar in both the dry and in water states. 
Previously\cite{ToBe} it is found that the wear rate for the NR compound in the dry state was nearly 
independent of the nominal contact pressure within the studied range (from $0.12$ to $0.43 \ {\rm MPa}$). 
This finding is confirmed by the results of the present study, in the pressure range from $0.16$ to $0.49 \ {\rm MPa}$. 

There are several effects that may result in different wear rates in the dry state compared to in water. 
First, even a small change in the friction coefficient can significantly affect the wear rate. 
This is predicted by the wear theory used above (see Ref. \cite{ToBe}) and has also been observed in the experiments of Rabinowicz, 
who found that the wear rate scales with the friction coefficient as $\sim \mu^5$ (see Fig. 3 in Ref. \cite{Rabi0}). 
This highlights the sensitivity of wear to the friction coefficient. 

However, in the present study, an opposite trend is observed for SBR: while the friction coefficient is lower in water, the wear rate is higher. This indicates that a different wear mechanism may be active in the presence of water. 

In the dry state, wear particles often remain trapped at the interface during reciprocal motion, where each sliding cycle involves multiple forward and backward passes over the same track. The presence of soft wear particles increases the contact area with the rubber block, which can lead to a higher friction coefficient even if the interfacial shear stress is reduced. A decrease in shear stress at the interface can reduce the wear rate, even when the contact area increases.

In contrast, under in-water conditions, these wear particles are likely carried away from the interface by the water flow. This can be observed in Figs. \ref{ALLwearCarbon.eps} and \ref{ALLwearSilicone.eps}, where wear particles tend to accumulate at the edges of the contact region. As a result, the contact interface remains cleaner, allowing higher localized shear stresses to develop in the asperity contact regions. These elevated shear stresses promote crack propagation, leading to increased wear rates.

\begin{figure}[tbp]
\includegraphics[width=0.4\textwidth,angle=0]{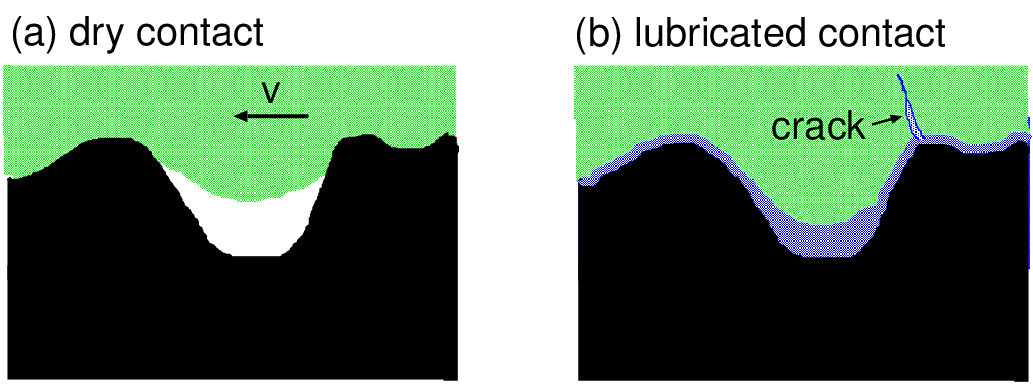}
\caption{\label{WATER.ReduceSherStress}
Rubber wear in water may be larger than that in the dry state due to reduced shear stress in the contact area, 
which allows more rubber to slip into cavities on the road surface. This picture assumes that the water in the cavity
is not sealed off by the surrounding rubber-substrate contact, but can leave the cavity, e.g. through roughness channels.
}
\end{figure}

Another explanation for this observation is the potential role of a thin ($\sim 1 \ {\rm nm}$) water film, 
which could reduce the frictional shear stress in the contact area and allow the rubber to penetrate deeper into surface cavities, 
thereby increasing the wear rate (see Fig. \ref{WATER.ReduceSherStress}), 
assuming that the road cavities are not filled with confined and pressurized water. 
This phenomenon is well known in a different context: 
it is easier to cut a rubber plate with a sharp knife if the interface is lubricated by water, soapy water, or another fluid.

Whether a thin fluid film exists at the interface depends not only on the contact pressure but also on the interfacial energies between 
rubber-water, rubber-countersurface, and water-countersurface. 
This involves dewetting and forced wetting transitions, which are influenced by the contact pressure and sliding speed, 
as discussed elsewhere \cite{wet1,wet2,wet3}.

In general, many experiments have demonstrated that water can influence crack propagation in polymers. For instance, in Polymethylmethacrylate (PMMA), the fracture toughness is higher in water than in air, with the work for fracture being approximately four times greater in water \cite{PMMA}. This phenomenon has been attributed to the interdiffusion of water into PMMA, which affects the mobility of the polymer chains and alters the crack-tip process zone. Specifically, the plastic zone at the crack tip remains a craze, but the length of fibrils at break increases, leading to higher toughness values.

Other factors that could influence crack propagation include capillary effects and stress corrosion. For example, the presence of water at the crack tip could modify the supply of oxygen or ozone, thereby affecting the rate of stress corrosion. Similar mechanisms may be important in rubber wear in water, depending on the chemical composition of the rubber and the presence of additives.

The relation between $\Delta x$ and $\gamma$ for the SBR and NR compounds, shown in Fig. \ref{1logGamma.2WearIntegrand.SBR.eps}, 
has the same qualitative form as observed in macroscopic experiments. In particular, the fatigue crack energies 
$\gamma_0 = 90$ and $66 \ {\rm J/m^2}$ used above for SBR and NR match those found in macroscopic experiments. 
This is expected because, in the limit of infinitesimally slow-moving cracks, viscoelastic energy dissipation vanishes, 
and other crack tip processes, such as cavitation, may also be negligible. However, this is not the case for larger $\gamma$, 
where macroscopic experiments reveal complex processes involving cavitation, long-range viscoelastic energy dissipation, 
and (for NR) strain crystallization. These processes may be strongly reduced at the micrometer length scale, 
resulting in much larger $\Delta x$ values than observed in macroscopic experiments. Indeed, the magnitude of 
$\Delta x$ required to explain the observed wear rate for SBR at $\gamma \approx 160 \ {\rm J/m^2}$ 
(where the wear rate integrand is maximal; see Fig. \ref{1logGamma.2WearIntegrand.SBR.eps}(b)) 
is a factor of $\sim 100$ larger than that observed in macroscopic experiments ($40 \ {\rm nm}$ instead of $0.4 \ {\rm nm}$). It has recently been suggested \cite{C1,C2} that very short cracks, which may prevail in the initial stage of the removal of a wear particle, may require a different approach than that used in the theory presented in Sec. 4. We are not aware of studies investigating the relationship between $\Delta x$ and $\gamma$ at the small length scales involved in mild rubber wear. Accurately determining this relationship represents an important research direction within the current wear theory framework.

Rubber wear results from the removal of small (micrometer-sized) rubber particles through crack propagation. When a smooth virgin rubber surface slides on a rough surface, a run-in period is required for a steady-state distribution of surface cracks to form. We have observed (not shown here) that at low nominal contact pressures ($\sim 0.1 \ {\rm MPa}$), a sliding distance of approximately $10 \ {\rm m}$ is needed to establish the steady-state distribution of surface cracks, resulting in a constant wear rate. This sliding distance corresponds to every point on the rubber surface being exposed to approximately 100 contacts with the road macroasperities, which is necessary to remove a rubber particle by fatigue crack propagation. This also implies that accurately characterizing the distribution of initial cracks on the rubber surface may be essential for reliable wear predictions.

\vskip 0.3cm
{\bf 7 Summary and conclusion}

In this study, we have investigated the wear behavior of styrene-butadiene rubber (SBR) and natural rubber (NR) 
on concrete surfaces under both dry and in-water conditions. Low sliding speeds ($\approx 3 \ {\rm mm/s}$) 
were used to minimize frictional heating and suppress hydrodynamic effects in water. 
The experimental results obtained under dry conditions show good agreement with predictions from a 
recently developed theory of rubber wear, offering insight into the fundamental wear mechanisms. 

For the two SBR compounds studied, significantly higher wear rates were 
observed in water compared to the dry state. The enhancement factors ranged from $1.5-2.5$ for the 
low-glass-transition compound ($T_{\rm g} \approx -50^\circ {\rm C}$) to approximately $4$ for 
the higher-glass-transition compound ($T_{\rm g} \approx -7^\circ {\rm C}$). For NR, almost no wear is 
observed in water at low nominal contact pressures ($\sigma_0 \approx 0.12$, $0.16$ and $0.25 \ {\rm MPa}$), 
while at $\sigma_0 \approx 0.36$ and $0.49 \ {\rm MPa}$, the wear rates are similar in both the dry and in-water states. 

The theory of rubber wear depends on the crack tip displacement $\Delta x$ in response to the time-dependent stress field from the road asperities. In this study, it is assumed that $\Delta x = 0$ for $\gamma < \gamma_0$ and $\Delta x = a (\gamma - \gamma_0)$ for $\gamma > \gamma_0$, which aligns with the expected relationship between the tearing energy $\gamma$ and $\Delta x$ when $\gamma$ is near the fatigue tearing energy $\gamma_0$. For $\gamma_0$, results from measurements on macroscopic samples are used. Consequently, the theory includes only one fitting parameter, $a$, chosen to reproduce the observed wear rates for a ``smooth'' concrete surface. Since the $\Delta x(\gamma)$ relationship is a material property, it should apply universally to any substrate. However, it is found that the amplitude parameter $a$ for micrometer-sized cracks might be up to $\sim 100$ times larger than observed for macroscopic cracks, indicating scale-dependent behavior not captured by traditional macroscopic measurements. Accurately determining the relationship between $\gamma$ and $\Delta x$ represents an important research direction within the current wear theory framework. Also, accurately characterizing the distribution of initial cracks on the rubber surface may be essential for reliable wear predictions.

\vskip 0.3cm
{\bf Author declarations:}
The authors have no conflicts to disclose.
All authors have contributed equally.

{\bf Data availability:}
The data that support the findings of this study are available
within the article. The data that support the findings of 
this study is available from the corresponding
author upon reasonable request.

\vskip 0.5cm
{\bf Appendix A: Rubber effective modulus}

\begin{figure}
\includegraphics[width=0.47\textwidth,angle=0.0]{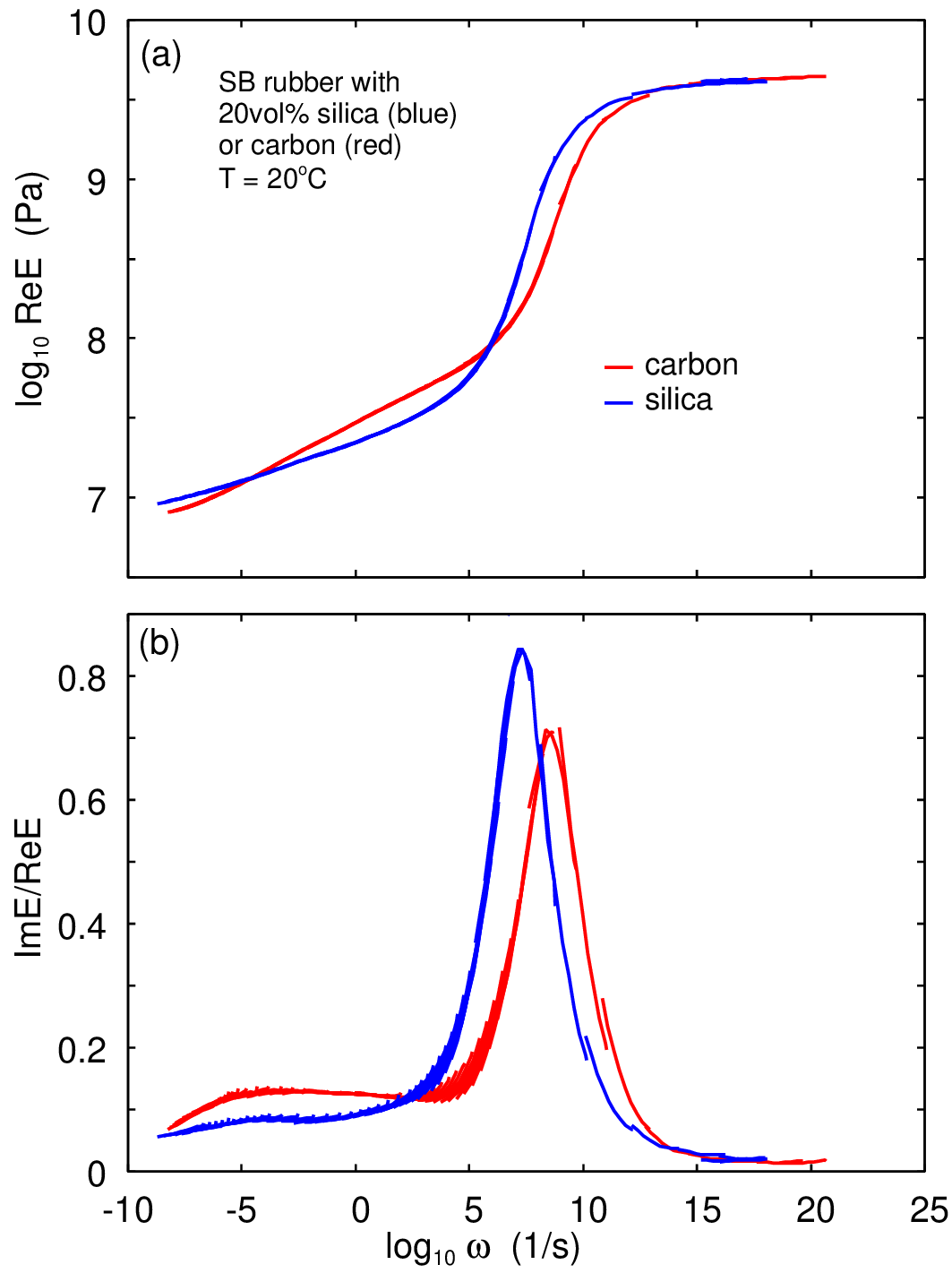}
\caption{
The dependency of the low strain ($\epsilon = 0.0004$) viscoelastic modulus
        on the frequency  $\omega$ for $T=20^\circ {\rm C}$ for SBR with 
	20vol\% carbon (red curves) and silica (blue). (a) shows the logarithm of the real part of the viscoelastic modulus
	and (b) the ratio ${\rm Im}E/{\rm E}$ as a function of the logarithm of the frequency.}
\label{1logOmega.2logReE.both.eps}
\end{figure}

The characteristic deformation frequency when a rubber block slides in contact with a road asperity is 
$\omega \approx v/r_0$, where $r_0$ is the linear size of the contact region. 
In the present case, the contact radius is $\approx 0.03 \ {\rm mm}$, and with a sliding speed of $v=3 \ {\rm mm/s}$, 
this gives deformation frequencies of approximately $100 \ {\rm s}^{-1}$. 
Fig. \ref{1logOmega.2logReE.both.eps} shows the dependency of the low-strain 
($\epsilon = 0.0004$) viscoelastic modulus on the frequency $\omega$ at $T=20^\circ {\rm C}$ 
for SBR with $20 \ {\rm vol}\%$ carbon (red curves) and silica (blue). 
(a) displays the logarithm of the real part of the viscoelastic modulus, and (b) the ratio ${\rm Im}E/{\rm E}$ as a 
function of the logarithm of the frequency.

\begin{figure}
\includegraphics[width=0.47\textwidth,angle=0.0]{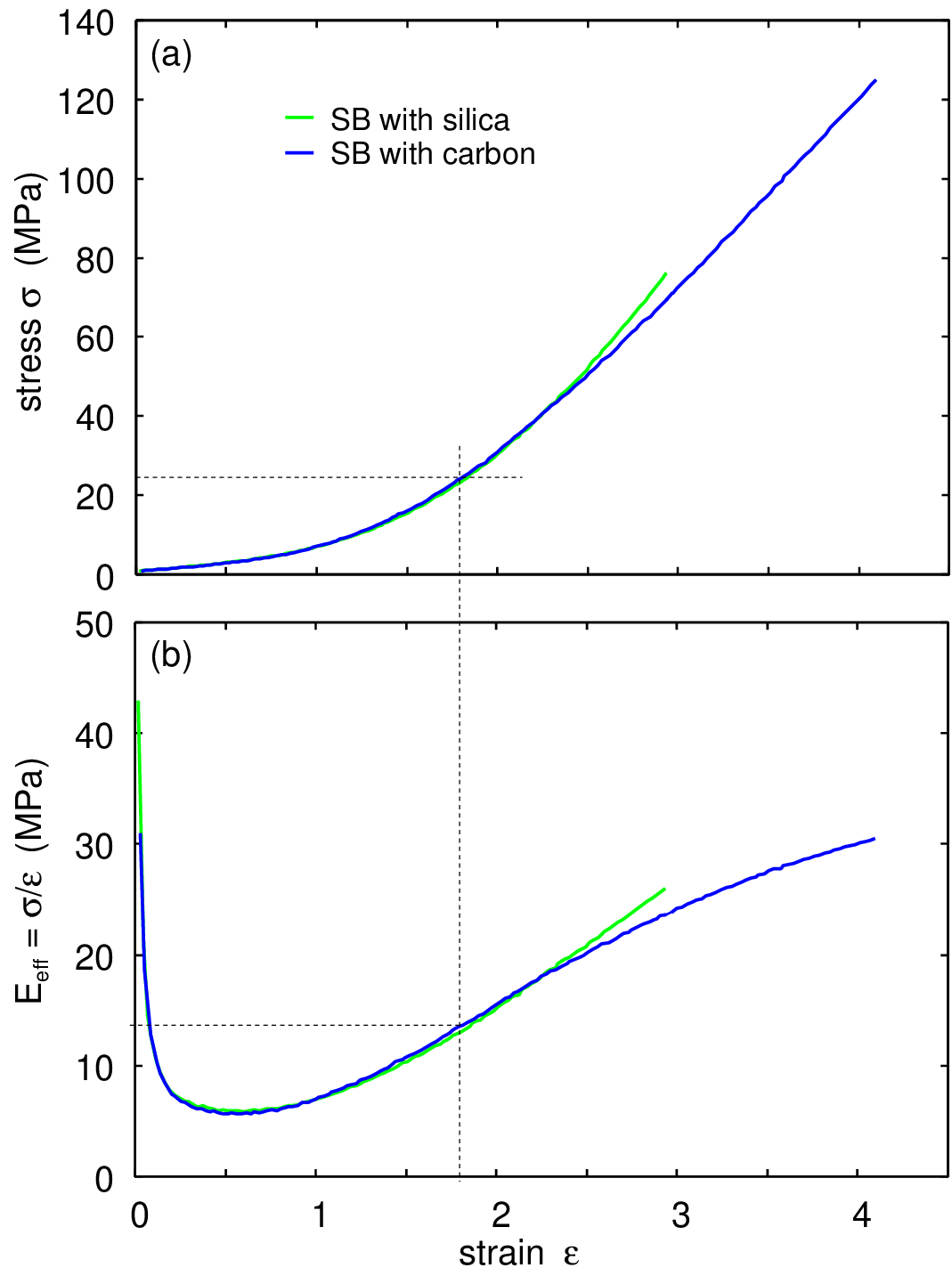}
\caption{
	The relation between the (physical) stress and the stain and (b)
	the effective (secant) modulus $E_{\rm eff} = \sigma/\epsilon$ for SBR filled with silica (green line) and with
	carbon black (blue line). The stress $\sigma$ 
	is the true (or physical) stress defined by $\sigma = F/A$, where $F$ is the elongation force and $A$
the rubber block cross section which depends on the strain. The strain is defined as usual $\epsilon = (L-L_0)/L_0$.
	The measurement was performed by elongation of a rubber strip at the strain rate $\dot \epsilon = 0.3 \ {\rm s}^{-1}$.}
\label{1strain.2stress.eps}
\end{figure}

The non-linear properties of the rubber (see Fig. \ref{1strain.2stress.eps}) have been studied at a strain rate 
$\dot \epsilon \approx 0.3 \ {\rm s}^{-1}$, corresponding to the characteristic frequency 
$\dot \epsilon /\epsilon \approx 0.2 \ {\rm s^{-1}}$ when the strain $\epsilon \approx 1.5$. 
When the deformation frequency increases from $0.2$ to approximately $100 \ {\rm s}^{-1}$, 
the low-strain modulus increases by a factor of $\sim 1.4$ for both compounds.

The stress-strain relation for filled rubbers is strongly non-linear. We define the effective (secant) modulus 
$E$ as the ratio between the (physical) stress and the strain, $E = \sigma /\epsilon$. Here, $\sigma = F/A$, 
where $F$ is the elongation force and $A$ is the cross-section of the rubber strip, which depends on the strain: 
$A = A_0/(1+\epsilon)$, where $A_0$ is the cross-section of the rubber strip before elongation. 
The strain $\epsilon$ is defined as $\epsilon = (L-L_0)/L_0$, where $L$ and $L_0$ are the lengths of the rubber strip 
after and before applying the force $F$, respectively. This secant modulus is required for the typical strain 
prevailing in the rubber-road asperity contact regions.

The stress in the asperity contact regions can be estimated using (4). With $E^* \approx 20 \ {\rm MPa}$, 
$\gamma \approx 200 \ {\rm J/m^2}$, and $r_0 \approx 13 \ {\rm \mu m}$, this gives $\sigma \approx 25 \ {\rm MPa}$. 
Fig. \ref{1strain.2stress.eps}(a) shows the stress-strain relation for the elongation of a strip of the NR used in this study. 
The stress $\sigma \approx 25 \ {\rm MPa}$ corresponds to a strain $\epsilon \approx 1.8$. 
The effective (secant) modulus $E$ for this strain is shown in Fig. \ref{1strain.2stress.eps}(b) 
and is approximately $E=14 \ {\rm MPa}$. Since the low-strain modulus is about $1.4$ times larger at 
a frequency of $100 \ {\rm s}^{-1}$ compared to $0.2 \ {\rm s}^{-1}$, we estimate that the relevant modulus 
is approximately $1.4 \times 14 \approx 20 \ {\rm MPa}$.

\vskip 0.5cm
{\bf Appendix B: Surface roughness power spectrum}

The most important quantity characterizing a rough surface is the surface roughness power spectrum \cite{Challange,Review}. 
The two-dimensional (2D) surface roughness power spectrum, $C({\bf q})$, which is central to the Persson contact mechanics theory, 
can be calculated from the height profile $z=h(x,y)$ measured over a square surface area. However, for surfaces with roughness having isotropic statistical properties, the 2D power spectrum can also be derived from the 1D power spectrum obtained from a line scan $z=h(x)$.

The 2D power spectrum is defined as \cite{JCPP,Preview}
$$C({\bf q}) = {1\over (2\pi )^2} \int d^2x \ \langle h({\bf x})h({\bf 0}) \rangle 
e^{-i{\bf q}\cdot {\bf x}}. \eqno(B1)$$
If the surface profile $z=h(x,y)$ of a two-dimensional surface area, is expressed as a sum (or integral) of plane waves:
$$h({\bf x}) = \int d^2q \ h({\bf q}) e^{i{\bf q}\cdot {\bf x}}, \eqno(B2)$$
then the 2D power spectrum can alternatively be written as:
$$C({\bf q}) = {(2 \pi )^2\over A_0} |h({\bf q})|^2, \eqno(B3)$$
where $A_0$ is the surface area. For surfaces with isotropic statistical properties, $C({\bf q})$ depends only 
on the magnitude $q=|{\bf q}|$ of the wavevector ${\bf q}$, where $q=2 \pi /\lambda$, with $\lambda$ being the wavelength of a surface roughness component.

Many surfaces, including the concrete surfaces studied here, exhibit a non-symmetric height distribution 
(i.e., not symmetric under $h \rightarrow -h$) \cite{ToBe}. For such surfaces, it is useful to study the top power spectrum $C_{\rm T}$, defined as \cite{Preview}
$$C_{\rm T} = {1\over (2\pi )^2} {A_0 \over A_{\rm T}} \int d^2x \ 
\langle h_{\rm T}({\bf x})h_{\rm T}({\bf 0}) \rangle 
e^{-i{\bf q}\cdot {\bf x}}, \eqno(B4)$$
where $h_{\rm T}({\bf x}) = h({\bf x})$ for $h > 0$ and zero otherwise, and $A_T/A_0$ is the fraction of the total (projected) area where $h > 0$. Similarly, the bottom power spectrum $C_{\rm B}$ can be defined using $h_{\rm B}({\bf x}) = h({\bf x})$ for $h < 0$ and zero otherwise.

The physical interpretation of $C_{\rm T}$ is that it represents the power spectrum of a surface where the roughness below the average plane is replaced by roughness with the same statistical properties as that above the average plane. This is the relevant power spectrum for theoretical calculations that assume random roughness, which ``looks the same'' above and below the average surface plane.

The wear experiments are conducted on two different concrete surfaces. The concrete blocks (pavers) are obtained from a ``Do-It-Yourself" shop in large quantities. Typically, each new wear experiment was performed on a fresh concrete block. These concrete blocks have been used in most of our earlier friction studies due to their stability (no or negligible wear of the concrete itself), and blocks obtained from the same batch have consistent nominal surface roughness. For each surface, at least three tracks are measured at different locations, each $25 \ {\rm mm}$ long.

Surface topography is measured using a Mitutoyo Portable Surface Roughness Measurement Surftest SJ-410 equipped with a diamond tip of radius $R=1 \ {\rm \mu m}$. The tip exerted a repulsive force of $F_N = 0.75 \ {\rm mN}$. The measurements had a step length (pixel) of $0.5 \ {\rm \mu m}$, a scan length of $L=25 \ {\rm mm}$, and a tip speed of $v=50 \ {\rm \mu m/s}$. 

\begin{figure}[h]
\includegraphics[width=0.95\columnwidth]{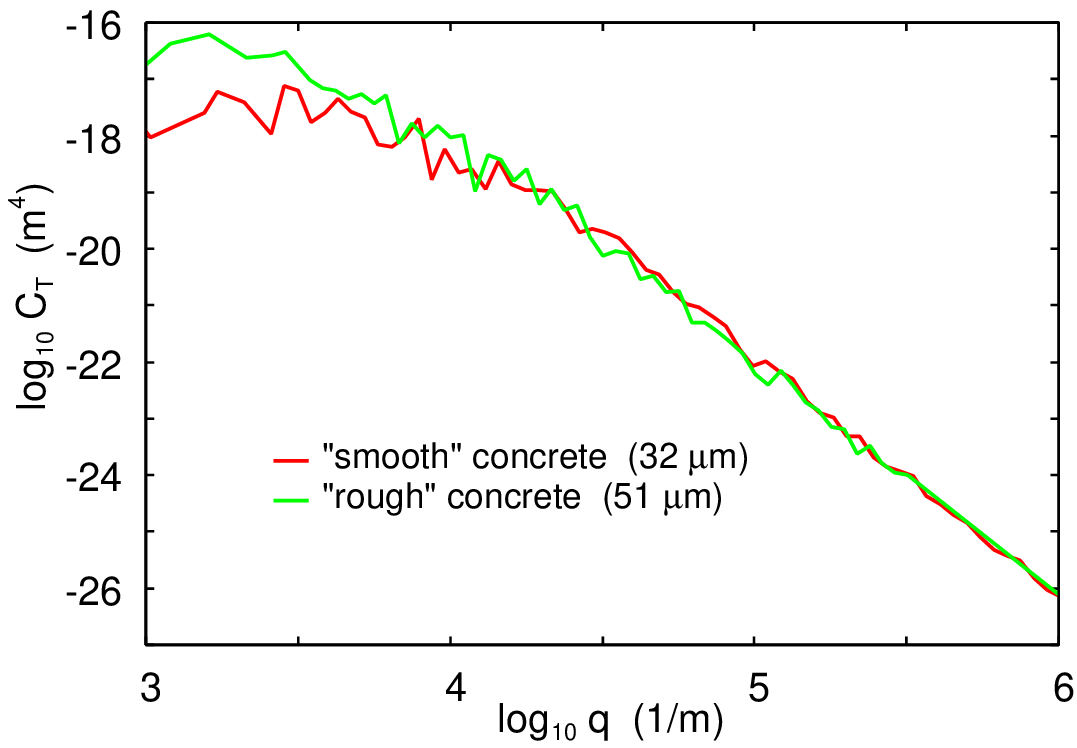}
\caption{\label{1logq.2logC.SMOOTH.ROUGH.FINAL.eps}
        The 2D top surface roughness power spectrum obtained from line scans on
the the ``smooth'' and ``rough'' concrete surface.
}
\end{figure}

In Fig. \ref{1logq.2logC.SMOOTH.ROUGH.FINAL.eps} we show the top power spectra of the smooth and rough concrete surfaces.
In each case, we have averaged over three line scan measurements. 
The rough and smooth concrete surfaces have the root-mean-square (rms) 
roughness amplitudes $51.3$ and $33.6 \ {\rm \mu m}$, respectively and the rms-slopes $0.46$ and $0.37$. 

\vskip 0.5cm
{\bf Appendix C: Elastic energies and the role of the normal stress on wear}

It is usually assumed that wear results mainly from the frictional shear stress, which is likely accurate as long as the friction is sufficiently high. However, in lubricated contact at low sliding speeds, the frictional shear stress may be relatively low even when asperity contact occurs. In such cases, the normal stress may significantly influence the wear rate. This topic relates to the extensive body of studies on crack formation in the indentation of brittle solids \cite{brittle,twocracks,constant,inden0}.

Rabinowicz presented experimental results supporting the importance of tangential stress in wear. Specifically, he demonstrated that for a wide range of dry and lubricated metals, the wear rate is proportional to $\mu^5$ for friction coefficients in the range $0.1 < \mu < 1$. However, this finding differs from studies on rubber wear, such as Ref. \cite{constant}, where wear rates are nearly independent of the friction coefficient from $\mu \approx 0.3$ to $\mu \approx 1$, followed by a very rapid increase in wear rates for $\mu > 1$ \cite{constant}. Regardless, the elastic energy resulting from the normal stress field will influence the wear rate to some extent.

\begin{figure}[h]
\includegraphics[width=0.99\columnwidth]{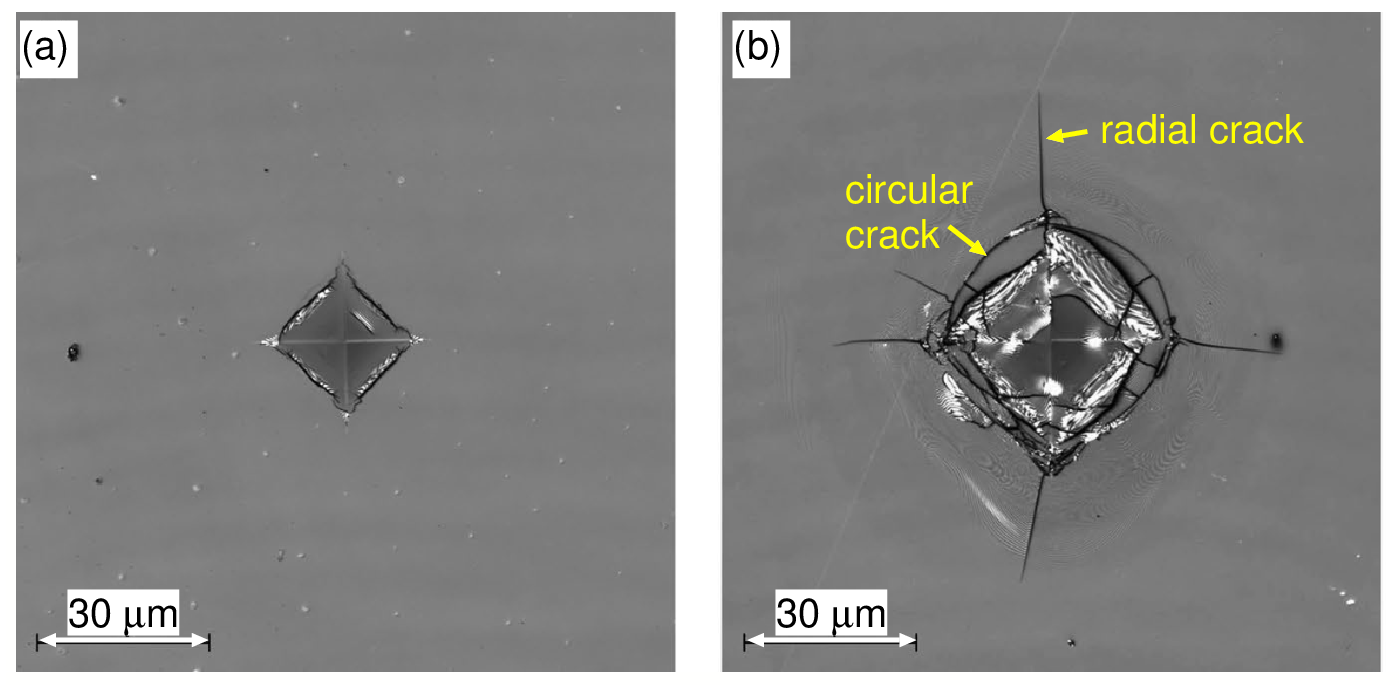}
\caption{\label{TwoIndent.eps}
Vickers indentation crack pattern in fused silica loaded with (a) 3 N and (b) 5 N. 
	Indents loaded with less than 3 N exhibit no crack propagation. Adapted from Ref. \cite{twocracks}.
	}
\end{figure}

Cracks at the surface of a solid can be induced by both normal and tangential stresses. The normal force primarily produces compressive deformations that cannot drive opening cracks, while tensile stresses outside the contact regions can generate circular or radial opening cracks if the elastic deformation field is sufficiently large [see Fig. \ref{TwoIndent.eps}(b)]. This effect can significantly impact the wear rate when the frictional shear stress is low.

Opening crack propagation has been observed in indentation experiments on brittle materials. Indentation-induced microfracture is common in brittle solids like silicates (quartz, fused silica) and ceramics \cite{brittle}. For a ``sharp'' indenter, a localized zone of irreversible (plastic) deformation forms near the contact point, providing nucleation centers for microcracks. Conversely, for a ``blunt'' indenter, contact conditions may remain entirely elastic until fracture onset. A classic example is the Hertzian cone fracture produced by indenting a flat surface with a relatively hard sphere. Initiation occurs from pre-existing surface flaws in the region of high tensile stress just outside the contact circle; the crack subsequently encircles the contact and propagates downward and outward.

Two main types of opening cracks typically propagate from the deformation zone: median vents, formed during indenter loading, extend downward below the contact point along planes of symmetry, and lateral vents, formed during unloading, spread sideways toward the specimen surface \cite{brittle}. 

Below a certain threshold loading, cracks are suppressed because insufficient stored elastic energy exists to drive crack propagation, as illustrated in Fig. \ref{TwoIndent.eps}. In silica glasses, the impression size at which this threshold occurs is typically on the order of tens of micrometers, which coincides with the typical size of wear particles. This correlation aligns with the expectation that crack propagation and wear occur only when sufficient stored elastic energy is present in the contact region.

The superposition of a tangential loading force, aside from altering the stress trajectory configurations, significantly enhances tensile stress, particularly at the trailing edge of the contact area. To account for the normal stress field in wear calculations (discussed in Sec. 3), we propose a simple approach: replacing the friction coefficient $\mu$ with an effective friction coefficient $\surd \mu^2+\mu_0^2$, where $\mu_0$ is a constant determined as follows.

Consider an elastic half-space $z>0$ and let ${\bf x} = (x,y)$ be a point in the surface plane.
If $\sigma_i = \sigma_{zi}$ denote the stress on the surface $z=0$ then the surface displacement 
$u_i ({\bf x})$ is linearly related to $\sigma_i({\bf x})$. The elastic energy
$$U={1\over 2} \int d^2x \ \sigma_i u_i = {1\over 2} \int d^2x \ (\sigma_x u_x +\sigma_z u_z)\eqno(C1)$$
where we have assumed $\sigma_y=0$. Using the theory of linear elasticity
$$u_x = {1\over 4 \pi G} \int d^2x' \ \bigg (\left [{2(1-\nu)\over |{\bf x}-{\bf x}'|} +2\nu {(x-x')^2 
\over |{\bf x}-{\bf x}'|^3}  \right ] \sigma_x ({\bf x}')$$
$$- (1-2\nu){(x-x')\over |{\bf x}-{\bf x}'|^2}\sigma_z ({\bf x}')\bigg )\eqno(C2)$$
and
$$u_z = {1\over 4 \pi G} \int d^2 x' \bigg [{2(1-\nu)\over |{\bf x}-{\bf x}'|}\sigma_z ({\bf x}') $$
$$+ (1-2\nu) {(x-x')\over |{\bf x}-{\bf x}'|^2} \sigma_x ({\bf x}')  \bigg ] \eqno(C3)$$
where $G=E/[2(1+\nu)]$ is the shear modulus.
Substituting (C2) and (C3) this in (C1) and assuming that $\sigma_i$ is constant for $r<R$ and zero for $r>R$
gives
$$U= {1\over 4 \pi G}  \int_{r,r' < R} d^2x d^2x' \ \bigg [ \left ( {1-\nu \over |{\bf x}-{\bf x}'|}
+ {\nu (x-x')^2 \over |{\bf x}-{\bf x}'|^3}\right ) \sigma_x^2$$  
$$+{1-\nu  \over |{\bf x}-{\bf x}'|}\sigma_z^2 \bigg ]$$ 
Since
$$\int_{r,r' < R} d^2x d^2x' \ {(x-x')^2 \over |{\bf x}-{\bf x}'|^3} = {1\over 2} \int_{r,r' < R} d^2x d^2x' \ {1 \over |{\bf x}-{\bf x}'|}$$
we get
$$U= {1\over 4 \pi G}  \int_{r,r' < R} d^2x d^2x' \ \bigg [  {1-\nu/2 \over |{\bf x}-{\bf x}'|}\sigma_x^2
+{1-\nu \over |{\bf x}-{\bf x}'|}\sigma_z^2 \bigg ]$$  
Writing $F_x = \pi R^2 \sigma_x$ and $F_z = \pi R^2 \sigma_z$ we get
$$U={1\over 2 K_x^*} F_x^2 + {1\over 2 K_z^*} F_z^2\eqno(C4)$$
where
$${1\over K_x^*} = {1\over 2 \pi G} {1\over \pi^2 R^4}  \int_{r,r' < R} d^2x d^2x' \  {1-\nu/2\over |{\bf x}-{\bf x}'|}$$
$${1\over K_z^*} = {1\over 2 \pi G} {1\over \pi^2 R^4}  \int_{r,r' < R} d^2x d^2x' \ {1-\nu \over |{\bf x}-{\bf x}'|}$$
We write ${\bf x} = R {\bf \xi}$ so that
$${1\over K_x^*} = {1\over 2 \pi G R} {1\over \pi^2 }  \int_{\xi,\xi' < 1} d^2\xi d^2\xi' \ {1-\nu/2\over |{\bf \xi}-{\bf \xi}'|}$$
$${1\over K_z^*} = {1\over 2 \pi G R} {1\over \pi^2}  \int_{\xi,\xi' < 1} d^2\xi d^2\xi' \ {1-\nu \over |{\bf \xi}-{\bf \xi}'|}$$
The integral
$$Q = {1\over \pi^2} \int_{\xi,\xi' < 1} d^2\xi d^2\xi' \ {1\over  |{\bf \xi}-{\bf \xi}'|} \approx 1.70$$
is easily performed by numerical integration. Using this definitions  and that $2G = E/(1+\nu)$ we get
$${1\over K_x^*} = {1\over \pi E^* R} Q {1-\nu/2 \over 1-\nu}\eqno(C5)$$
$${1\over K_z^*} = {1\over \pi E^* R}Q\eqno(C6)$$
where $E^* = E/(1-\nu^2)$. For rubber-like materials $\nu \approx 0.5$ so that
$${1\over K_x^*} \approx {2.55\over \pi E^* R}, \ \ \ \ {1\over K_z^*} \approx {1.70\over \pi E^* R}$$
If $F_x/F_z = 1$ (corresponding to the friction coefficient $\mu=1$), $U_x/U_z \approx 2.55/1.70 = 1.5$
and more energy will be stored in the deformation field induced by the tangential stress but the energies 
are of similar magnitude. Since a large part of the elastic energy $U_z$ is compressive it cannot be used,
at least in the initial state of a crack propagation, to remove wear particles. That is, the elastic energy
release rate derived from $U_z$ may be smaller than from $U_x$. We suggest to use when estimating the wear rate
the effective elastic energy $U_{\rm el} = U_x+\epsilon U_z$ with $\epsilon < 1$, instead of $U_{\rm el}$ given by (1).
Using (C4)-(C6) with $F_x = \mu F_z$ this gives
$$U_{\rm el} = {F_z^2\over 2 K_x^*} \left (\mu^2 + \epsilon {K_x^*\over K_z^*}\right )
={F_z^2\over 2 K_x^*} \left (\mu^2 + \mu_0^2 \right )\eqno(C7)$$
where
$$\mu_0^2 = {2 \epsilon (1-\nu) \over 2-\nu }$$
Using the elastic energy (C7) instead of (1) is equivalent to replacing $\mu$ in (4) with
$\surd (\mu^2 + \mu_0^2)$.

\end{document}